\documentclass[showpacs,amsfonts,preprint,aps,prc]{revtex4} 

\usepackage{graphicx}
\usepackage{bm}
\usepackage{epsfig}
\newcommand\NPA{Nucl. Phys. A} 
\newcommand\NPB{Nucl. Phys. B}

\newcommand\PLB{Phys. Lett. B} 
\newcommand\PR{Phys. Rep.} 
 
\newcommand\PRL{Phys. Rev. Lett.} 
\newcommand\PRC{Phys. Rev. C} 
\newcommand\PRD{Phys. Rev. D} 
\newcommand\PRV{Phys. Rev.}

\newfam\BMath 
\font\BMathL=cmmib10  
\font\BMathl=cmmib7 
\font\BMathm=cmmib5 
\textfont\BMath=\BMathL \scriptfont\BMath=\BMathl 
\scriptscriptfont\BMath=\BMathm 
 
\renewcommand\P{{\fam\BMath p}}

\renewcommand\a{\alpha} 
\renewcommand\b{\beta} 
 
\renewcommand\d{\delta} 
\newcommand\df{\delta^{(4)}}

\newcommand\m{\mu} 
\newcommand\n{\nu} 
 
\newcommand\p{\pi} 
\newcommand\s{\sigma}

\renewcommand\L{\Lambda}      % Often used as a large scale or baryon 
\newcommand\cc{{\cal C}}

\newcommand\cm{{\cal M}}   % probability amplitude 
 
\newcommand\cp{{\cal P}}   % principal value for example 

\newcommand\ra{\rightarrow} 
 
\newcommand\lra{\leftrightarrow}  
 
\newcommand\Lra{\longrightarrow}

\newcommand{\nonum}{\nonumber} 
 
\newcommand{\half}{\textstyle{\frac{1}{2}}}

\newcommand\be{\begin{equation}} 
\newcommand\ee{\end{equation}} 
\newcommand\bea{\begin{eqnarray}} 
\newcommand\eea{\end{eqnarray}} 
\newcommand\beal{\begin{align}}
\newcommand\eeal{\end{align}}
 
\newcommand\ba{\begin{array}} 
\newcommand\ea{\end{array}} 
\newcommand\bc{\begin{center}}
\newcommand\ec{\end{center}}

\newcommand\eref[1]{Eq.~(\ref{#1})} 
\newcommand\etwref[2]{Eqs.~(\ref{#1}) and (\ref{#2})} 
\newcommand\ethref[3]{Eqs.~(\ref{#1}), (\ref{#2}) and (\ref{#3})} 
\newcommand\eforef[4]{Eqs.~(\ref{#1}), (\ref{#2}), (\ref{#3}) and (\ref{#4})}

\newcommand\bfi{\begin{figure}} 
\newcommand\efi{\end{figure}} 
\newcommand\bpi[1]{\begin{picture}#1} 
\newcommand\epi{\end{picture}} 
 
\newcommand{\fref}[1]{Fig.~\ref{#1}}

\def\jou#1#2#3#4{{#1} {\bf #2}, #3 (#4)} 
 
\newcommand\intp{\frac{d^3p}{(2\pi)^3 2 E_p}}

\newcommand\ptv{\P_\perp} 
 
\newcommand\tBE{{\text{BE}}}

\newcommand\xmn{{x_{\min}}} 
\newcommand\xmx{{x_{\max}}}  
\newcommand\zmn{{z_{\min}}} 
\newcommand\zmx{{z_{\max}}} 
 
\newcommand\F{\Phi}  
 
\newcommand\pll{\parallel}  
\newcommand\plln{\!\!\parallel\!\!}  
 
\newcommand\ing{\frac{1}{\n_g}}  

\newcommand\tri{\triangle} 

\newcommand\ben{\begin{displaymath}} 
\newcommand\een{\end{displaymath}} 
\newcommand\bean{\begin{eqnarray*}} 
\newcommand\eean{\end{eqnarray*}} 

\begin{document} 
 
%\begin{flushright} 
%\footnotesize \sffamily 
%\end{flushright} 

%\draft 
%{ 
%\wideabs{ 
 
\title{
Out-of-Equilibrium Collinear Enhanced Equilibration in the Bottom-Up 
Thermalization Scenario in Heavy Ion Collisions
}

\author{S.M.H. Wong} 
 
\affiliation{Department of Physics, The Ohio State University, 
Columbus, OH 43210} 

%\date{Version 2 -- \today} 
%\date{Version 2 -- January 18, 2004} 
 
%\maketitle 
 
\begin{abstract}  

Experimental measurement of the elliptic flow parameter $v_2$ and
hydrodynamic model together showed that thermalization in the
central region at the Relativistic Heavy Ion Collider to be perplexingly
fast. This is a mystery in itself since none of the numerical
perturbative QCD models are able to achieve such a feat. 
By exploiting a theoretical oversight on collinear processes in 
an out-of-equilibrium system it is argued that, in the bottom-up 
thermalization scenario, equilibration can proceed at a higher rate 
than what is expected in the conventional perturbative QCD picture.

\end{abstract} 

%\vspace{1.0cm} 
%\pacs{PACS: 25.75.-q, 12.38.Bx, 12.38.Mh, 24.85.+p}  
%}  
 
\maketitle

%%%%%%%%%%%%%%%%%%%%%%%%%%%%%%%%%%%%%%%%%%%%%%%%%%%%%%%%%%%%%%%%%%%%%%%%%%%%%%%
\section{Introduction} 
\label{s:intro} 
%%%%%%%%%%%%%%%%%%%%%%%%%%%%%%%%%%%%%%%%%%%%%%%%%%%%%%%%%%%%%%%%%%%%%%%%%%%%%%%

Recently measurement by STAR \cite{st1,st2,rs}, PHENIX \cite{lacey} 
and PHOBOS \cite{park} collaborations of the elliptic flow parameter 
$v_2$ in the central region at the Relativistic Heavy Ion Collider 
(RHIC) at Brookhaven agreed with that generated from simulation using 
the hydrodynamic model \cite{o92,ksh,khh,hkhrv}. As is well-known 
that hydrodynamical model required that a system to be in 
complete equilibrium. This implies that the central region in the 
collisions at RHIC achieved equilibrium within an extremely short 
time of around 0.6 fm/c. In case one has any doubts about this
conclusion, the authors investigated an alternate scenario where the
collision system was thermalized only in the transverse directions 
\cite{hzwo}. We find that it is impossible to achieve the same $v_2$ 
results while keeping the same final transverse momentum distribution.    
Even trying to maintain the latter required very drastic modification 
to the initial conditions so that these become unrealistic. This is
both good and bad news from a theoretical point of view. It is good
because a fully thermalized environment at least in the central
region permits a lot of simplifications in the calculation. 
A thermalized system is much simpler than  an out-of-equilibrium 
system. However this is also bad news because it shows that we have 
not fully understood what really happens in the initial stage of the 
collisions. There are two reasons for that. First it was shown in 
\cite{mg1,mg2,mg3} that the special case of the parton cascade model 
\cite{kg1,kg2,pcm} with only elastic collisions required either an 
initial parton density or the elastic cross section 15 times greater 
than the actual value in order to reproduce the same elliptic flow 
measured in the experiments. This is a huge factor, one does not 
expect, based on today's knowledge, that including inelastic 
collisions will produce a large enough total cross section. Second 
it is our opinion that none of the parton based numerical models 
for simulating heavy ion collisions, parton cascade included, is 
showing such perplexingly fast thermalization \cite{kg1,kg2,pcm,wong1}.  
They do show signs that the system is approaching equilibrium 
at a rapid pace even in the face of the longitudinal expansion,  
but none of these models managed to thermalize within 
as short an interval as 0.6 fm/c. We do not consider the results 
from other non-parton based models because they use other less 
well understood and less rigorous mechanisms such as strings, hadronic 
collisions etc or the combinations any of these. In any case it
is hard to justify the use of any hadronic mechanisms in view of the 
energy density that can be reached at RHIC energies. There is one 
other mechanism that can be at work which is that as the system 
cools, the average energy of the system must come down. This
will lead to an progressive increase in the strength of the 
interactions \cite{wong2}. This mechanism can certainly help with 
equilibration. However this mechanism does not come into play until 
at a much later stage in the collisions, well beyond the 0.6 fm/c 
limit imposed by the $v_2$ result.  
 
To solve this very rapid equilibration problem, the first question 
that one has to decide is what mechanism leads to this 
fantastic result. Is it perturbative or nonperturbative in 
nature? Parton cascade model is supposed to have incorporated
all known and relevant perturbative QCD effects \cite{pcm}. One 
can easily conclude that the as-yet-unknown mechanism must be 
nonperturbative. Nevertheless another picture of the nuclei
exists that involved viewing the initial nuclei as a frozen 
crystalline glass of QCD color \cite{cgc}. Initial conditions 
produced from this model have been used to investigate and tried to 
answer the perturbative or nonperturbative equilibration question 
\cite{am}. This line of investigation resulted eventually in ref. 
\cite{betal}, this paper used a result already pointed out in 
\cite{wong1} which is that thermalization cannot come from leading 
order processes alone, higher order processes are important too. 
It was concluded in \cite{betal} that the equilibration 
could proceed perturbatively. However this paper did not address 
how rapid a perturbative QCD based equilibration 
could proceed. Even if equilibration in heavy ion collisions can 
be achieved via perturbative QCD alone, can it be done with
such phenomenal speed as revealed by the combination of elliptic 
flow measurements and the hydrodynamic model? There is a clear
advantage of a perturbation theory based rapid equilibration as 
opposed to a nonperturbative one. The former is well understood 
theoretically and tested in experiments while the latter is 
less rigorous and based on a much less solid foundation. 
Therefore before one draws any definitive conclusion based on the 
results of the parton based models and gives up on perturbative QCD, 
another careful examination is warranted. 

We will show in this paper that there is indeed a new mechanism 
that has been overlooked largely because of a piece of well-known 
knowledge that has been overly generalized and assumed to hold 
under all circumstances even though no careful study has been done
in some situations. By this we means the cancellation of collinear
divergence for a theory of massless particles which has been shown 
to be true in the vacuum \cite{kin,ln} and in an equilibrium system. 
Indeed this is true in the vacuum as total cross-sections are free 
of any collinear logarithms as these are cancelled between the real 
and virtual contributions. This is also true for a system of massless 
particles in equilibrium. Because of this, one tends to assume 
tacitly that the same will also hold in a non-equilibrium environment. 
{\em This is not true. In general this cancellation between real and 
virtual graphs ceases to occur for a system that is not in equilibrium.} 
One might be alarmed by this statement since there should not be any 
remaining divergences. This is reasonable but thanks to screening in a 
medium such divergences do not exist. The cancellation of collinear 
logarithms for a system in equilibrium is thus a double safeguard 
against this kind of divergences. In such a system, removing either 
screening or the cancellation of collinear logarithms alone does 
not result in any divergences. In the rest of the paper, we will 
show in details how there are necessarily non-cancelled collinear 
logarithms in an out-of-equilibrium parton plasma and how this 
result can be exploited to bring about a thermalization via 
perturbative QCD that can proceed at a higher speed than any parton 
models have shown thus far.

%%%%%%%%%%%%%%%%%%%%%%%%%%%%%%%%%%%%%%%%%%%%%%%%%%%%%%%%%%%%%%%%%%%%%%%%%%%%%%%
\section{The most relevant collision processes for equilibration}
\label{s:rel_coll}
%%%%%%%%%%%%%%%%%%%%%%%%%%%%%%%%%%%%%%%%%%%%%%%%%%%%%%%%%%%%%%%%%%%%%%%%%%%%%%%

In the early stage of the system, the plasma is gluon dominated 
because of the small-x growth in the gluon distribution in the 
nucleons and also because of the stronger gluon-gluon interactions 
than that of gluon-(anti)quark or (anti)quark-(anti)quark. Therefore
for our present attempt at explaining the very fast thermalization 
seen in experiments, one only has to consider a pure gluon plasma. 
In general there are many possible interactions even only among 
gluons. Not all of them are useful for the purpose of equilibration. 
We will group the different contributions below and discuss
the importance and relevance of them with regard to thermalization. 

%%%%%%%%%%%%%%%%%%%%%%%%%%%%%%%%%%%%%%%%%%%%%%%%%%%%%%%%%%%%%%%%%%%%%%%%%%%%
\subsection{$2 \longleftrightarrow 2$ scattering}
%%%%%%%%%%%%%%%%%%%%%%%%%%%%%%%%%%%%%%%%%%%%%%%%%%%%%%%%%%%%%%%%%%%%%%%%%%%%

\begin{itemize} 

\item[(i)]{Small angle collision: 
At the leading order, it is the simple 2-to-2 gluon-gluon scattering. 
It is well known that this process is dominated by small angle 
scatterings because of the familiar Coulomb exchange 
divergence at small momentum transfer. Although the probability
of this process is large, the outgoing momenta are not that different
from the incoming ones since they are only rotated somewhat in momentum
space from the incoming momenta. Thus this normally dominant process
is in fact terribly inefficient at driving the system towards
equilibration. 
}

\item[(ii)]{Large angle collision: 
At the same order, there is the remaining large angle scattering
process. The outgoing momenta are necessarily very different 
from the incoming ones therefore large angle scatterings are
much better at redistributing momenta which is the essence of
what thermalization is about. The drawback of this process is that 
the probability is not nearly as large as the small angle collisions. 
There will be an occasional large angle collision amongst the
much more frequent small angle collisions. 
}

\end{itemize}

At this level of the basic 2-to-2 scattering, the possibility for
the driving force behind the equilibration process is restricted 
between a probable but inefficient small angle scatterings and 
a less probable but more efficient large angle collisions. This
is not very promising. Indeed as have been pointed out in the 
second reference of \cite{wong1}, the elastic scattering process
is not the main driving force for equilibration. This is because 
in that numerical study, the entropy generation rate from 
2-to-2 scatterings is not only comparable but sometimes even 
subdominant to that of the 2-to-3 gluon processes. This point
of view is reinforced in a later paper \cite{betal}. Therefore 
higher order processes must be considered. The key question
is which higher processes?

%%%%%%%%%%%%%%%%%%%%%%%%%%%%%%%%%%%%%%%%%%%%%%%%%%%%%%%%%%%%%%%%%%%%%%%%%%%%
\subsection{$2 \longleftrightarrow 3$ scattering}
%%%%%%%%%%%%%%%%%%%%%%%%%%%%%%%%%%%%%%%%%%%%%%%%%%%%%%%%%%%%%%%%%%%%%%%%%%%%

We now attempt to identify what processes at this order are
most relevant and most important for thermalization. 

\begin{itemize} 

\item[(i)]{Small angle collisions with a small angle gluon 
           radiation/absorption:
At order $\a_s^3$ there is the possibility of a gluon emission 
on top of the main scattering. Since the emitted gluon can be 
of any magnitude limited only by energy-momentum conservation,
this is a much more flexible process for achieving equilibration.
However flexibility is not the only factor that one has to consider. 
In this case, all the momenta are aligned around the directions 
of the two incoming momenta. So there is still no significant 
momentum transfer here. Besides there is the extra power of 
$\a_s$ that reduces the effectiveness of this contribution. The
same applies to the small angle absorption.  
}

\item[(ii)]{Small angle collisions with a large angle gluon 
            radiation/absorption: 
This type of contributions come in two different forms. One can
be reclassified as large angle collisions with a small angle emission
or absorption to be discussed below. Those contributions that 
cannot be reclassified are as before down by a power of $\a_s$
although in view of the large angle gluon emission and the
small angle scattering, they are potentially useful for 
achieving equilibration. 
}

\item[(iii)]{Large angle collisions with a small angle gluon 
            radiation/absorption:
Again large angle collisions are important for sizable momentum
rearrangement as already mentioned, what is more important 
is the well-known fact that small angle gluon emission or absorption
accompanying a hard collision can give rise to a large collinear
logarithm in the probability provided that the collision is 
sufficently hard. In the medium this is determined by both the
hardness of the collisions and the medium screening, this will be
seen in later sections. These kind of large logarithms can compensate
for the small power of $\a_s$ that comes with the gluon emission
or absorption. Thus the contribution from this type of processes is 
comparable in size to that from the leading order. 
}

\item[(iv)]{Large angle collisions with a large angle gluon 
            radiation/absorption:
Large angle processes sadly do not receive any collinear 
logarithmic compensation of the small coupling. Processes discussed
in (iii) are definitely more important for this reason. 
}

\end{itemize}

%%%%%%%%%%%%%%%%%%%%%%%%%%%%%%%%%%%%%%%%%%%%%%%%%%%%%%%%%%%%%%%%%%%%%%%%%%%%
\subsection{Higher order processes} 
%%%%%%%%%%%%%%%%%%%%%%%%%%%%%%%%%%%%%%%%%%%%%%%%%%%%%%%%%%%%%%%%%%%%%%%%%%%%

Going to even higher orders, one can have various mixtures of 
large and small angle processes given that the possibilities are
nearly endless as one goes to higher and higher order. 
Afterall we are using perturbative QCD, higher processes are 
in general less important. From the above discussion, we have
ascertained that large angle processes are useful for equilibration
because of their ability to rearrange the momentum distribution. 
However they are weighed down by the coupling and have no 
singularity to enhance the frequency of their occurrence. It is 
certain therefore that there cannot be too many large angle 
subprocesses in any given scattering and there cannot be 
none either. The simplest way to proceed is to combine both large 
and small angle processes. Since there must be at least 
one large angle process and the coupling associated with small 
angle subprocesses can only be compensated by a large collinear 
logarithm if there is a hard scattering and/or small screening, it 
is logical that for contribution at any order to have one hard 
collision to be accompanied by all sorts of collinear subprocesses. 
These type of contributions are now comparable in size to that 
of the leading order 2-to-2 scattering. These are the processes
that we considered to be most important for equilibration and 
we will concentrate on these contributions and consider them further 
up to order $\a_s^3$ in the rest of this paper. {\em Note that we 
are not relying solely on the momentum rearranging power of the large 
angle collisions. They serves mainly to direct and guide where the 
remaining gluon radiations will populate the momentum space. It is 
these radiations that actually help to bring about thermalization 
of the system.}. In a way we are arguing for using multi-gluon 
processes to achieve rapid thermalization not dissimilar to what
was done in \cite{shx} for achieving chemical equilibration. 
Whereas in \cite{shx} they included mostly every $n g \lra m g$
process indiscriminately, we consider only what in our opinion
the most important processes for thermalization here. 

Notwithstanding of what was mentioned in the introduction, any 
alert readers will rightly object that one cannot use collinear
processes in this way because of the cancellation between the
real and virtual contribution \cite{kin}. This is true
only if there is a sum over degenerate states \cite{ln}. The 
equilibration of a gluon plasma, and in fact for other many-body
systems as well, is not at all inclusive because momentum
regions are separate from one another. The processes cannot be 
any more exclusive than this but it is usual to study  
interactions and momentum states collectively, in which case 
one has indeed to consider this cancellation. As mentioned 
in the introduction and we will see below, this cancellation does 
not hold in an out-of-equilibrium system \cite{wip}. But there is 
no need to worry about any divergence because where there is screening, 
the interaction probability is always finite.

%%%%%%%%%%%%%%%%%%%%%%%%%%%%%%%%%%%%%%%%%%%%%%%%%%%%%%%%%%%%%%%%%%%%%%%%%%%%%%%
\section{The leading and next-to-leading order collision processes} 
\label{s:l-&-ntl}
%%%%%%%%%%%%%%%%%%%%%%%%%%%%%%%%%%%%%%%%%%%%%%%%%%%%%%%%%%%%%%%%%%%%%%%%%%%%%%%

An out of equilibrium system is necessarily time dependent, there 
must be a transport equation to study such a system. The simplest 
and possibly the most used is the Bolzmann equation 
 \be  
    p^\m \frac{\partial f(x,p)}{\partial x^\m} = C(x,p)   \, , 
 \label{eq:tr} 
 \ee 
On the right hand side are the collision terms which contain 
all the microscopic interactions. Since they are the main reason
that a many-body system can equilibrate, we will spend the
rest of the article studying these. As is commonplace in 
perturbative QCD, the collinear gluons are either on-shell or 
are nearly on-shell so that only the transverse degrees of freedom
are relevant. Therefore a physical gauge such as the lightcone
gauge will be used throughout. 

%%%%%%%%%%%%%%%%%%%%%%%%%%%%%%%%%%%%%%%%%%%%%%%%%%%%%%%%%%%%%%%%%%%%%%%%%%%%
\subsection{The leading gluon-gluon collision process}
%%%%%%%%%%%%%%%%%%%%%%%%%%%%%%%%%%%%%%%%%%%%%%%%%%%%%%%%%%%%%%%%%%%%%%%%%%%%

\bfi
\epsfig{figure=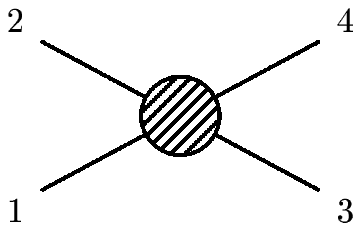,width=4.0cm}
\caption{The basic scattering of gluon 1 with gluon 2 
resulting in gluon 3 and 4. Likewise one can view this equally as
gluon 3 scatters with gluon 4 giving gluon 1 and 2.} 
\label{f:2to2} 
\efi

The basic gluon-gluon collision is well-known. Graphically we 
represent it by \fref{f:2to2}. The collision term is 
 \bea 
    C_{12\lra 34}(p_1) &=& -\int [dp_2][dp_3][dp_4] D_{12,34} \,
         \frac{1}{2!}\overline{|\cm_{12\ra 34}|^2}  
 \nonum \\ 
    & & \hspace{1cm} \times  
        \bigl[f_1 f_2 (1{+}f_3)(1{+}f_4)-(1{+}f_1)(1{+}f_2) f_3 f_4\bigr] \,.
 \label{eq:c22} 
 \eea 
Here $f_j=f(p_j)$, $[dp]$ represents the invariant momentum-space
integration measure for an on-shell gluon with momentum $p$,  
 \be
   [dp] = {d^4p\over (2\pi)^3}\, \theta(p^0)\, \delta(p^2) = \intp\,,  
 \ee 
with $E_p=|\bm{p}|$ for massless gluons, and $D_{12,34}$ is the 
energy-momentum conserving $\d$-function 
 \be
    D_{12,34} = (2\p)^4 \df(p_1+p_2-p_3-p_4)   \; . 
 \label{eq:delta} 
 \ee 
The square of the matrix element is given by \cite{ckr} 
 \be 
  \overline{|\cm_{12\ra 34}|^2} = 
  \ing \sum |\cm_{12\ra 34}|^2 = 72\,\n_g \p^2 \a_s^2(Q^2)   
    \Big (3-\frac{ut}{s^2} -\frac{us}{t^2} -\frac{st}{u^2} \Big )  
 \ee  
where the sum is over spins and colors, $s=(p_1{+}p_2)^2$, 
$t=(p_1{-}p_3)^2$, and $u=(p_1{-}p_4)^2=-s-t$. 
For the scale of the running coupling constant $\a_s(Q^2)$ we can take
$Q^2=\min(|s|,|t|,|u|)$. At momentum scales where there are three active 
quark flavors, it can be parametrized as  
 \be   
    \a_s (Q^2) =\frac{4\p}{\b_0 \ln (Q^2/\L_{(3)}^2)}  
 \ee 
where $\L_{(3)}=0.246$\,GeV and $\b_0 =9$. We will assume that $Q$ is  
significantly larger than the scale $\L$ associated with  
non-perturbative QCD effects.

%%%%%%%%%%%%%%%%%%%%%%%%%%%%%%%%%%%%%%%%%%%%%%%%%%%%%%%%%%%%%%%%%%%%%%%%%%%%
\subsection{The gluon-gluon collision with collinear emission or absorption}
%%%%%%%%%%%%%%%%%%%%%%%%%%%%%%%%%%%%%%%%%%%%%%%%%%%%%%%%%%%%%%%%%%%%%%%%%%%%

For 5-gluon processes there are two distinct contributions to the 
collision terms: 
 \bea 
   C_{12\lra 345}(p_1) &=& - \frac{1}{3!} \int [dp_2] d\F_{345} 
   D_{12,345}\, \overline{|\cm_{12\ra 345}|^2}        
 \nonum \\  
     && \hspace{1cm} \times  
        \bigl[f_1 f_2 (1{+}f_3)(1{+}f_4)(1{+}f_5) 
              -(1{+}f_1)(1{+}f_2) f_3 f_4 f_5 \bigr], 
 \label{eq:c23} 
 \eea   
and 
 \bea
   C_{123\lra 45}(p_1) &=& - \frac{1}{2!2!} \int d\F_{23} d\F_{45} 
   D_{123,45}\, \overline{|\cm_{123\ra 45}|^2}
 \nonum \\   
     && \hspace{1cm} \times    
   \bigl[f_1 f_2 f_3 (1{+}f_4)(1{+}f_5)
         -(1{+}f_1)(1{+}f_2)(1{+}f_3) f_4 f_5 \bigr] .  
 \label{eq:c32}  
 \eea 
We have introduced a compact notation for the multiparticle phase 
space: 
 \be
    \int d\F_{12\dots m} = [dp_1][dp_2]\cdots [dp_m]    \, . 
 \ee 
Every such factor is accompanied by a symmetry factor $1/m!$.
The energy-momentum conserving delta functions $D_{12,345}$ and 
$D_{123,45}$ are the obvious generalizations of \eref{eq:delta}. 
 
Suppose the 5-gluon processes $12\lra 345$ in \eref{eq:c23} and  
$123\lra 45$ in \eref{eq:c32} consist of a hard-scattering collision  
involving four gluons. Then two of the incoming or outgoing gluons  
can be hard but collinear. If the two gluons ($i$,$j$) are both  
in the initial or both in the final state, collinear means that their  
invariant mass $s_{ij}=(p_i+p_j)^2$ is significantly less than $Q^2$,  
where $Q$ is the momentum scale of the hard scattering. If one is in the  
initial state and one is in the final state, collinear means that their  
momentum transfer $|t_{ij}|=-(p_i-p_j)^2$ is likewise significantly less than  
$Q^2$. For the process $12\lra 345$, there are nine possibilities for the  
collinear pair of gluons ($i$,$j$). By exploiting the Bose symmetry, we can  
take the collinear pair to be (1,5), (2,5) or (4,5). For the process  
$123\ra 45$, there are again nine possibilities for the collinear pair. Again  
by exploiting the Bose symmetry, we can take the collinear pair to be  
(1,3), (2,3) or (4,3). For each of the possible collinear pairs,  
we wish to calculate the contribution to the collision terms $C(p)$  
from the collinear region of phase space. We will label the contribution  
in which gluons $i$ and $j$ are collinear by $i\!\pll\! j$.

%%%%%%%%%%%%%%%%%%%%%%%%%%%%%%%%%%%%%%%%%%%%%%%%%%%%%%%%%%%%%%%%%%%%%%%% 
\subsubsection{Both collinear gluons in the initial or final state} 
\label{ss:bif} 
%%%%%%%%%%%%%%%%%%%%%%%%%%%%%%%%%%%%%%%%%%%%%%%%%%%%%%%%%%%%%%%%%%%%%%%% 

We first consider the case in which the collinear gluons are either  
both in the initial state or else both in the final state and neither  
is the external gluon 1 (of course there is nothing special about
gluon 1 except its momentum has been chosen to be the argument of the
collision terms). We begin by considering the scattering process  
$12\ra 3ij$ in which the two collinear gluons are both in the final  
state with momenta $p_i$ and $p_j$. The matrix element $\cm_{12\ra 3ij}$  
has a pole in the invariant mass $s_{ij}$. In the collinear region  
in which $s_{ij} \ll Q^2$, we can interpret the process as 
proceeding through a hard-scattering $12\ra 34^*$ that creates 
a virtual gluon $4^*$ with momentum $p_{4^*}=p_i+p_j$ followed by the 
decay of the virtual gluon into the two collinear gluons $i$ and $j$. 
 
It is convenient to introduce light-cone variables defined by the  
direction of the 3-momentum $\P_{4^*}=\P_i+\P_j$ in the rest frame of  
the plasma. The momenta of the collinear gluons and the virtual gluons 
can be written as  
\bea p_i &=& \Big (   z  p^+_4 , \frac{\ptv^2}{z p^+_4},      \ptv \Big )  \\  
     p_j &=& \Big ((1-z) p^+_4 , \frac{\ptv^2}{(1-z) p^+_4}, -\ptv \Big )  \\ 
 p_{4^*} &=& \Big ( p^+_4 ,\frac{s_{ij}}{p^+_4} ,\mathbf{0_\perp}  \Big )   
\eea 
where $p^+_4$ is the total light-cone momentum of the collinear  
gluons. The variable $z$ is the fraction of that light-cone momentum  
that comes from gluon $i$. The light-cone energies are determined  
by the mass-shell constraints $p_i^2=p_j^2=0$. The invariant mass  
$s_{ij}$ of the virtual gluon is determined by conservation of  
light-cone energy:  
\be   s_{ij} = \frac{\ptv^2}{z} + \frac{\ptv^2}{1-z}  
      = \frac{\ptv^2}{z(1-z)}   \; .   
\label{eq:sij} 
\ee  
The phase space integral in light-cone variables is 
\be  [dp] = \frac{dp^+}{4\p p^+} \frac{d^2 \ptv}{(2\p)^2}   
          = \frac{dz}{4\p z} \frac{d^2 \ptv}{(2\p)^2}  \;, 
\ee  
the latter form applies when $p$ is the momentum of the daughter  
of a parent gluon from which a fraction $z$ of its light-cone  
momentum has been inherited. We define the momentum $p_4$ of an on-shell  
gluon that has the same light-cone momentum as the virtual gluon $4^*$ but  
whose light-cone energy is determined by the mass shell constraint  
$p^2_4=0$:  
\be p_4 = \Big ( p^+_4 ,0, \mathbf{0_\perp} \Big ) \;. 
\ee  
 
We proceed to make approximations to the collision integral  
\eref{eq:c23} (with 45 replaced by $ij$) that are valid in the  
collinear region. In the amplitude for $12\ra 34^*$, we can 
replace the momentum $p_{4^*}$ of the virtual gluon by the quasi 
on-shell momentum $p_4$, since the invariant mass 
$s_{ij}$ is small compared to the hard-scattering scale.  
In the amplitude for the decay $4^*\ra ij$, we keep only the most 
singular term which will give rise to a logarithm of $Q$ after 
integrating over phase space. The square of the matrix element, 
averaged over initial spins and colors and summed over those of the 
final state, then reduces to 
\be \sum |\cm_{12\ra 3ij}|^2 \simeq \frac{8\p \a_s(s_{ij})}{s_{ij}}  
    P_{gg}(z) \sum |\cm_{12\ra34}|^2  
\label{eq:rme1} 
\ee 
where $P_{gg}(z)$ is the unregularized $g\ra gg$ splitting function  
\be P_{gg}(z) = 2\, C_A \Big ( \frac{1-z}{z}+\frac{z}{1-z}+z(1-z) \Big )  
\label{eq:Pgg}  
\ee 
where $C_A=3$ is a color factor. The scale of the explicit factor of 
$\a_s$ in \eref{eq:rme1} is $\sqrt{s_{ij}}$.  
We can make similar approximations in the phase space integrals. 
We first express the phase space for gluons $i$ and $j$ in an 
iterated form involving an integral over the momentum of the virtual 
gluon $4^*$  
\be \int [dp_i] [dp_j] D_{12,3ij} = \int \frac{ds_{ij}}{2\p}  
    [dp_{4^*}] D_{12,34^*} \int [dp_i] [dp_j] D_{4^*,ij}  \; .   
\label{eq:ps}  
\ee 
In the collinear region, we can approximate $p_{4^*}$ by $p_4$ in the  
measure $[dp_{4^*}]$ and in the delta function $D_{12\ra 34^*}$. 
The expression for the phase space integral \eref{eq:ps} then 
reduces to  
\be  \int [dp_i] [dp_j] D_{12,3ij} \simeq  
     \int [dp_4] D_{12,34} \frac{dz\; d\ptv^2}{16\p^2 z(1-z)}  
   = \int [dp_4] D_{12,34} \frac{dz\; d s_{ij}}{16\p^2} \; .  
\label{eq:rps1}  
\ee 
Combining the approximations \etwref{eq:rme1}{eq:rps1} and using  
\eref{eq:sij}, we get    
\be \int d\F_{3ij} D_{12,3ij} \sum |\cm_{12\ra 3ij}|^2 \simeq  
    \int d\F_{34} D_{12,34} \sum |\cm_{12\ra 34}|^2 \times \int  
    \frac{dz\; ds_{ij}}{s_{ij}}  \frac{\a_s(s_{ij})}{2\p} P_{gg}(z)   \; . 
\ee  
 
\bfi
\epsfig{file=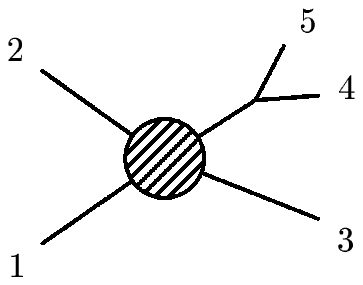, width=4.0cm}
\caption{An example of the 2-to-3 or 3-to-2 gluon process when 
gluon $4\plln 5$.}
\label{f:2to3f}
\efi

Our final step is to use a collinear approximation in the distribution  
functions $f_i$ and $f_j$ for the gluons. Since their momenta have  
components of order $Q$, we can neglect their transverse momenta 
and approximate them by $f_4(z)$ and $f_4(1-z)$, where 
\be  f_i(z) \equiv f(z p_i)   \; .  
\ee  
Thus our final expression for the contribution to the collision 
integral \eref{eq:c23} from the region in which gluons 4 and 5 are 
collinear is  
\bea C^{4\pll 5}_{12\lra 345}(p_1) & \simeq & -\ing \int [dp_2] 
     d\F_{34} D_{12,34} \frac{1}{3!} \sum |\cm_{12\ra 34}|^2  
     \times \int dz \frac{ds_{45}}{s_{45}}  \frac{\a_s(s_{45})}{2\p}  
     P_{gg}(z)                                                   \nonum \\  
     && \times \Big [ f_1 f_2 (1+f_3)(1+f_4(z))(1+f_4(1-z)) 
                     -(1+f_1)(1+f_2) f_3 f_4(z) f_4(1-z) \Big ] \; .  
                                                                 \nonum \\  
\label{eq:bif-23}  
\eea  
Depicted in \fref{f:2to3f} is the feynman graph for this contribution
when $4\plln 5$. The contribution from the collinear regions $3\plln 4$ 
and $3\plln 5$ are given by exactly the same expression after appropriate 
relabelling of the momenta. 
 
The contribution to the collision integral \eref{eq:c32} from the  
collinear region $2\plln 3$ is given by a similar expression. After  
relabelling the dummy momentum variables $45\ra 34$, it can be  
written as  
\bea C^{2\pll 3}_{123\lra 45}(p_1) & \simeq & -\ing \int [dp_2] d\F_{34}  
     D_{12,34} \frac{1}{2!2!} \sum |\cm_{12\ra 34}|^2   
     \times \int dz \frac{ds_{23}}{s_{23}}  \frac{\a_s(s_{23})}{2\p}  
     P_{gg}(z)                                                 \nonum \\ 
   \!& \times &\! \Big [ f_1 f_2(z) f_2(1-z) (1+f_3)(1+f_4) 
                     -(1+f_1)(1+f_2(z))(1+f_2(1-z)) f_3 f_4 \Big ] \; .   
                                                               \nonum \\    
\label{eq:bif-32} 
\eea  
In both \etwref{eq:bif-23}{eq:bif-32}, the ranges of the integrals 
over $z$ and $s_{ij}$ have not been specified. The ranges in which 
our collinear approximation remain valid will be specified later after 
a discussion of hard-thermal loop corrections.

%%%%%%%%%%%%%%%%%%%%%%%%%%%%%%%%%%%%%%%%%%%%%%%%%%%%%%%%%%%%%%%%%%%%%%%%% 
\subsubsection{One collinear gluon in the initial and the other  
in the final state} 
\label{ss:oif} 
%%%%%%%%%%%%%%%%%%%%%%%%%%%%%%%%%%%%%%%%%%%%%%%%%%%%%%%%%%%%%%%%%%%%%%%%% 

We next consider the case in which one of the collinear gluons is in the 
initial state and one is in the final state and neither is the external 
gluon 1. We begin by considering the scattering process $1i\ra 34j$, where  
gluons $i$ and $j$ are collinear. The matrix element $\cm_{1i\ra 34j}$ has  
a pole in the variable $t_{ij}=(p_i-p_j)^2$. In the collinear region  
$|t_{ij}| \gg Q^2$, we can interpret the process as the splitting of 
gluon $i$ into a real gluon $j$ and a virtual gluon $2^*$ with  
momentum $p_{2^*}=p_i-p_j$, followed by the hard scattering  
$12^*\ra 34$. 
 
We introduce light-cone variables defined by the direction of  
$\P_i$ in the rest frame of the plasma. The momenta of 
the collinear gluons and the virtual gluon $2^*$ can be written as 
\bea p_i &=& \Big ( \frac{p^+_2}{x}, 0, \mathbf{0_\perp}  \Big )          \\  
     p_j &=& \Big ( \frac{(1-x) p^+_2}{x}, \frac{x \ptv^2}{(1-x) p^+_2}, 
                   -\ptv \Big )                                           \\  
 p_{2^*} &=& \Big ( p^+_2, \frac{t_{ij}+\ptv^2}{p^+_2}, \ptv  \Big )   
\eea 
where $p^+_2$ is the difference between the light-cone momenta of 
the collinear gluons. The variable $x$ is the fraction of the light-cone  
momentum of gluon $i$ that is carried by the virtual gluon $2^*$. The  
light-cone energies of $i$ and $j$ are determined by the mass shell  
constraints $p^2_i =p^2_j=0$. The invariant mass of the virtual gluon is  
determined by the conservation of light-cone energy  
\be  t_{ij} = -\frac{\ptv^2}{1-x}  \; .  
\label{eq:tij} 
\ee 
We define the momentum $p_2$ of an on-shell gluon that has the same 
light-cone momentum as the virtual gluon $2^*$ and the same transverse 
momentum as gluon $i$:  
\be  p_2 = (p^+_2, 0, \mathbf{0_\perp} )  \; .   
\ee 
 
\bfi
\epsfig{file=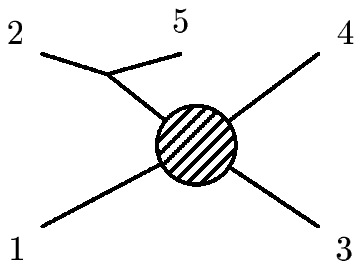, width=4.0cm}
\caption{An example of the 2-to-3 or 3-to-2 gluon process when 
gluon $2\plln 5$.}
\label{f:2to3i}
\efi

We proceed to make approximations to the collision integral 
\eref{eq:c23} (with the pair (2,5) replaced by ($i$,$j$)) that are 
valid in the collinear region \fref{f:2to3i}. The square of the matrix 
element can be approximated as follows 
\be \sum |\cm_{1i\ra 34j}|^2 \simeq \frac{8\p \a_s(|t_{ij}|)}{|t_{ij}|}  
    P_{gg}(x) \sum |\cm_{12\ra34}|^2  
\label{eq:rme2}  
\ee 
where the scale of the explicit factor of $\a_s$ is $\sqrt{-t_{ij}}$. 
The phase space integral can be approximated by  
\be  \int [dp_i] [dp_j] D_{1i,34j} \simeq  
     \int [dp_2] D_{12,34} \frac{dx\; d\ptv^2}{16\p^2 (1-x)} 
   = \int [dp_2] D_{12,34} \frac{dx\; d (-t_{ij})}{16\p^2} \; .   
\label{eq:rps2}  
\ee 
The distribution functions $f_i$ and $f_j$ can be approximated by 
$f_2({1 \over x})$ and $f_2({1-x \over x})$, respectively.  
Combining all these approximations, the final expression for the 
contribution to the collision integral \eref{eq:c23} from the region  
$2\plln 5$ is  
\bea C^{2\pll 5}_{12\lra 345}(p_1) & \simeq & -\ing \int [dp_2] d\F_{34}  
     D_{12,34} \frac{1}{3!} \sum |\cm_{12\ra 34}|^2                     
         \times \int dx \frac{dt_{25}}{t_{25}} \frac{\a_s(|t_{25}|)}{2\p}  
         P_{gg}(x)                                                 \nonum \\ 
     & & \times \Big [ f_1 f_2(\mbox{$1 \over x$}) (1+f_3)(1+f_4)  
                       (1+f_2(\mbox{$1-x \over x$}))  
                      -(1+f_1)(1+f_2(\mbox{$1 \over x$})) f_3 f_4 
                        f_2(\mbox{$1-x \over x$}) \Big ] \;.       \nonum \\  
\label{eq:oif-23} 
\eea  
The identical expression is obtained for $2\plln 3$ or $2\plln 4$.  
 
We can derive a similar expression for the contribution to the collision  
integral \eref{eq:c32} from regions in which one of the collinear 
gluon is in the initial state, the other is in the final state, and 
neither one is $p_1$. The contribution from the region $3\plln 5$  
can be written as after relabelling of momenta  
\bea C^{3\pll 5}_{123\lra 45}(p_1) & \simeq & -\ing \int [dp_2] d\F_{34}  
     D_{12,34} \frac{1}{2!2!} \sum |\cm_{12\ra 34}|^2                    
        \times \int dx \frac{dt_{35}}{t_{35}} \frac{\a_s(|t_{35}|)}{2\p}  
        P_{gg}(x)                                                   \nonum \\  
     & & \times \Big [ f_1 f_2 (1+f_3) f_4(\mbox{$1-x \over x$})   
                       (1+f_4(\mbox{$1 \over x$}))   
                      -(1+f_1)(1+f_2) f_3 (1+f_4(\mbox{$1-x \over x$}))  
                          f_4(\mbox{$1 \over x$}) \Big ] \; .   \nonum \\ 
\label{eq:oif-32}  
\eea 
An identical expression is obtained, after relabelling, for the regions 
$2\plln 4$, $2\plln 5$, and $3\plln 4$. In \etwref{eq:oif-23}{eq:oif-32}, 
the ranges of $x$ and $t_{ij}$ in which our collinear approximations 
remain valid will be specified later after a discussion of  
hard-thermal-loop corrections.

%%%%%%%%%%%%%%%%%%%%%%%%%%%%%%%%%%%%%%%%%%%%%%%%%%%%%%%%%%%%%%%%%%%%%%%%%%%%
\subsection{Hard-thermal loop corrections}  
\label{s:htl} 
%%%%%%%%%%%%%%%%%%%%%%%%%%%%%%%%%%%%%%%%%%%%%%%%%%%%%%%%%%%%%%%%%%%%%%%%%%%%
 
One important class of higher order corrections to the 4-gluon collision 
integral $C_{12\ra 34}(p_1)$ is hard-thermal-loop (HTL) corrections to 
the propagators of the four hard scattered gluons \cite{bp1,bp2}. These  
corrections arise from hard gluons in the medium that interact with one of  
the hard-scattered gluons but are then scattered back into their original  
momentum state. These forward scattering contributions must be added  
coherently over the distribution of hard gluons in the medium. The effect of  
these HTL corrections on transverse gluons with momenta near the light cone  
is very simple. They become quasi-particles with a mass $m_t$ given by 
\be m^2_t = \frac{24 \a_s(\m^2)}{\p} \int^\infty_0 dp p f(p)  
\ee 
where $f(p)$ is the distribution of hard gluons in the medium. An 
appropriate choice for the scale $\m$ of $\a_s$ is the momentum scale that  
dominates the integral. As an estimate for the scale, we will use 
\be  \m^2 = \frac{\int^\infty_0 dp p^3 f(p)}{\int^\infty_0 dp p f(p)} 
     \; . 
\label{eq:mt} 
\ee 
In the case of the equilibrium Bose-Einstein distribution, this description  
gives $\m=\sqrt{2/5} \p T$. We will usually refer to the transverse gluon  
quasi-particle simply as gluons and to the mass given by \eref{eq:mt} as 
the transverse mass. The transverse mass is related to the Debye screening 
mass $m_D$ simply by $m_t=m_D/\sqrt{2}$ \cite{wel}. This relation arises from  
the structure of HTL corrections to the gluon propagator which are strongly 
constrained by gauge invariance.  
 
The HTL corrections will have particularly important effects in the regions 
of phase space where gluons become collinear. In particular, they 
provide infrared cutoff on the integrals over the momentum fraction and 
invariant masses that appear in the collinear contributions to the 
collision integrals. We will use the HTL corrections to estimate the 
boundaries of the integration regions where our collinear approximations 
remain valid.  
 
We first consider the integral over $z$ and $s_{45}$ in the $4\plln 5$ 
contribution to the collision integral $C_{12\ra 345}$ which is given 
in \eref{eq:bif-23}. One effect of the HTL corrections is to replace 
the propagator $1/s$ of the virtual gluon by $1/(s-m_t^2)$. Another 
effect is to change the mass-shell constraints on the collinear gluons 
to $p^2_4=p^2_5=m_t^2$. This changes the expression \eref{eq:sij} for 
the invariant mass of the virtual gluon to  
\be  s_{45} = \frac{m_t^2+\ptv^2}{z(1-z)}   \; . 
\ee 
For fixed $z$, the lower limit on $s_{45}$ occurs at $\ptv =0$: 
\be  s_{\min} = \frac{m_t^2}{z(1-z)}        \; . 
\label{eq:smn} 
\ee 
The upper limit on $s$ is set by the scale $Q^2$ of the hard 
scattering. We choose to write it as  
\be  s_{\max} = c^2 Q^2 + m_t^2    
\label{eq:smx} 
\ee 
where $c$ is a constant near 1. The change from varying $c$ by a  
factor of 2 should be included in the theoretical error. We will 
take the virtuality $s_{45}-m_t^2$ of the virtual gluon to also 
provide the scale of the factor $\a_s$ in the collinear correction 
factors. The integral over $s_{45}$ can then be evaluated analytically 
\be \int^{s_{\max}}_{s_{\min}} \frac{ds}{s-m_t^2} \frac{\a_s(s-m_t^2)}{2\p}  
   = \frac{2}{\b_0} \ln \left (\frac{\ln\Big (c^2 Q^2/\L^2\Big )} 
     {\ln\Big ((1-z+z^2)m_t^2/z(1-z)\L^2 \Big )} \right )   \; .  
\label{eq:ln1}
\ee 
The condition $s_{max} > s_{min}$ constrains the integration region 
for $z$ to the range $z_{\max} > z > z_{\min}$, where   
\bea  z_{\max} &=& \half \Big (1+\sqrt{1-4m_t^2/(c^2 Q^2+m^2_t)} \Big ) \\  
\label{eq:zmn} 
      z_{\min} &=& \half \Big (1-\sqrt{1-4m_t^2/(c^2 Q^2+m^2_t)} \Big ) \; .  
\label{eq:zmx} 
\eea  
The condition $z_{\max} > z_{\min}$ also imposes a lower bound on $Q^2$: 
\be   Q^2 > \frac{3}{c^2} m^2_t    \; .   
\ee 
 
We next consider the integral over $x$ and $t_{25}$ in the $2\plln 5$ 
contribution to the collision integral $C_{12\ra 345}$, which is given 
in \eref{eq:oif-23}. The HTL corrections replace the propagator 
$1/t_{25}$ of the the virtual gluon by $1/(t_{25}-m_t^2)$. They  
also change the mass shell constraint to $p^2_2=p^2_5=m_t^2$. The 
resulting expression \eref{eq:tij} for the invariant mass of the  
virtual gluon is replaced by  
\be t_{25} = -\frac{m_t^2+\ptv^2}{1-x}   \; . 
\ee 
For fixed $x$, the lower limit on $-t_{25}$ occurs at $\ptv=0$: 
\be   -t_{\min} = \frac{m_t^2}{1-x}    \; , 
\ee 
the upper limit on $-t_{45}$ is set by the scale $Q$ of the  
hard scattering. We choose to write it as  
\be   -t_{\max} = c^2 Q^2 -m_t^2    
\ee  
where $c$ is again a constant near 1. The change from varying $c$ by 
a factor of 2 should be included in the theoretical error. We will 
take the virtuality $-t_{25}+m_t^2$ of the virtual gluon to also be 
the scale of $\a_s$ in the collinear correction factor. The integral 
over $t_{25}$ can then be evaluated analytically 
\be \int^{-t_{\max}}_{-t_{\min}} \frac{d(-t)}{-t+m_t^2}  
    \frac{\a_s(-t+m_t^2)}{2\p} 
  = \frac{2}{\b_0} \ln \left ( \frac{\ln\Big (c^2 Q^2/\L^2\Big )}  
     {\ln\Big ((2-x) m_t^2/(1-x)\L^2 \Big )} \right )   \; .  
\label{eq:ln2}
\ee  
The condition $-t_{\max} > -t_{\min}$ constrains the integration range  
for $x$ to the region $x<x_{\max}$, where  
\be  \xmx = 1- \frac{m_t^2}{c^2 Q^2}    \; . 
\ee 
For the lower limit, we will set 
\be  \xmn = \frac{m_t^2}{c^2 Q^2}       \; .  
\ee 
The condition $x_{\max} >0$ imposes a lower bound on $Q^2$ 
\be  Q^2 > \frac{1}{c^2} m_t^2    \; . 
\ee 
Note that in the limit where $Q^2/m_t^2 \gg 1$, there is little
distinction between the different limits since $\zmx \simeq \xmx$
and $\zmn \simeq \xmn$.

%%%%%%%%%%%%%%%%%%%%%%%%%%%%%%%%%%%%%%%%%%%%%%%%%%%%%%%%%%%%%%%%%%%%%%%%%%%%
\subsection{The collinear virtual contributions at the next-to-leading order}
\label{s:virt} 
%%%%%%%%%%%%%%%%%%%%%%%%%%%%%%%%%%%%%%%%%%%%%%%%%%%%%%%%%%%%%%%%%%%%%%%%%%%%
 
We have not considered virtual corrections to the basic hard 
scattering discussed in Sec. \ref{s:l-&-ntl}. These are of the 
same order in $\a_s$ as the 5-gluon processes. In the collinear 
limit, these corrections can be written in a compact way.  
To see this, recall that in the vacuum theory if one starts out with 
a virtual gluon, it has a chance of splitting into two gluons with 
momentum fractions $z$ and $1-z$. The associated variation in the 
probability density is \cite{ap} 
\be  \cp_{gg} + d\cp_{gg} = \d(1-z) + \frac{\a_s}{2\p} P^+_{gg}(z) ds 
\label{eq:prob-T0} 
\ee 
while the gluon reduces the square of its invariant mass from $s$ 
to a value between $s$ and $s-ds$ for a timelike branching. For  
spacelike branching, one substitutes $dt$ for $ds$ and $t$ is changed 
to a value between $t$ and $t+dt$. $P^+_{gg}(z)$ is the usual kernel 
of the Dokshitzer-Gribov-Lipatov-Altarelli-Parisi (DGLAP) 
equation \cite{ap,gl,dsz} and it carries also a $\d(1-z)$ function 
part. It is the `plus' distribution regularized gluon splitting 
function \cite{ap,esw}. Since momentum is always conserved, the total 
momentum must always be equal to that of the original virtual 
gluon independent of whether splitting occurred, then $P^+_{gg}(z)$ 
must satisfy the identity (in the absence of $q$ and $\bar q$)  
\be \int^1_0 d z z P^+_{gg}(z) = 0        
\ee 
or if we write out $P^+_{gg}(z)$ separately the parts with   
and without the $\d(1-z)$ function  
\be  P^+_{gg}(z) = P_{gg}(z)|_+ + \cc_{T=0}\; \d(1-z)  \; , 
\ee 
where $\cc_{T=0}$ is the constant in front of the $\d$-function  
part of $P^+_{gg}(z)$, momentum conservation gives   
\be \int^1_0 d z \Big \{ z P_{gg}(z) |_+ 
   +\cc_{T=0}\; \d(1-z) \Big \} = 0   \; .   
\label{eq:virt-T0} 
\ee  
The $\d$-function part has its origin from the virtual correction to  
the splitting in the collinear limit. $P_{gg}(z) |_+$ is the same 
as our $P_{gg}(z)$ defined in \eref{eq:Pgg} except the 
$1/(1-z)$ part has been regularized by the `plus'-distribution  
prescription and so is replaced by $1/(1-z)_+$. One can therefore 
write in a compact way the correction as   
\be  \cc_{T=0} = -\int^1_0 dz  z P_{gg}(z) |_+ 
               = -\frac{1}{2} \int^1_0 dz  P_{gg}(z) |_+   \; .   
\label{eq:ct0} 
\ee 
This is a clever way to deduce the virtual correction from 
momentum conservation \cite{ap,esw}. The last expression on the 
right is obtained by using the property $P_{gg}(z) = P_{gg}(1-z)$ 
which follows also from the fact that in every branching two gluons 
are created for every one that gets destroyed. 

For gluon splitting in a medium because of the accompanying particle 
distributions, the virtual corrections cannot be deduced so simply
especially when the system is out-of-equilibrium. However since the
basic gluon splitting remains the same, one can reasonably expect 
that only the integrand of \eref{eq:ct0} will be modified by some
factor involving the particle distributions of particle 
participating in the splitting. That is we expect
\be  \cc_{T} = -\frac{1}{2} \int^1_0 dz  P_{gg}(z) |_+ H(f, z)  \;, 
\label{eq:ct} 
\ee 
where $H$ is a function of the particle distributions and the
momentum fraction $z$, which will be seen presently. 

Because of the medium as discussed in Sec. \ref{s:htl}, HTL 
corrections naturally establish limits on the $z$ integration within 
which the collinear approximation holds. These limits suffice as 
a regularization, we need not use the plus-distribution prescription. 
As a result the splitting function will always be written as $P_{gg}(z)$
from now on.

%%%%%%%%%%%%%%%%%%%%%%%%%%%%%%%%%%%%%%%%%%%%%%%%%%%%%%%%%%%%%%%%%%%%%%%% 
\subsubsection{Type I: virtual correction with vacuum contribution}
\label{ss:vc1} 
%%%%%%%%%%%%%%%%%%%%%%%%%%%%%%%%%%%%%%%%%%%%%%%%%%%%%%%%%%%%%%%%%%%%%%%% 

\bfi
\epsfig{file=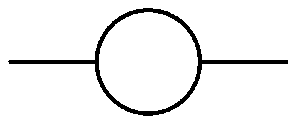, width=4.0cm}
\caption{The self-energy insertion that gives the relevant medium 
virtual contributions.} 
\label{f:se}
\efi

We divide the virtual corrections into two types, one that has a 
contribution in the vacuum and the other has not. Bearing in mind 
that for every real process beyond the leading order, there is a 
virtual contribution that can be considered as its partner. They are 
related simply by originating from two different cuts of the 
same diagram \cite{kin}. For this reason the pair of real and 
virtual contribution are closely related especially in the 
collinear limit. This facts allow them to be derived with some ease. 
The complete correction is a medium self-energy insertion shown in 
\fref{f:se} but only the type I part of this will be discussed 
in this section.

\bfi
\epsfig{file=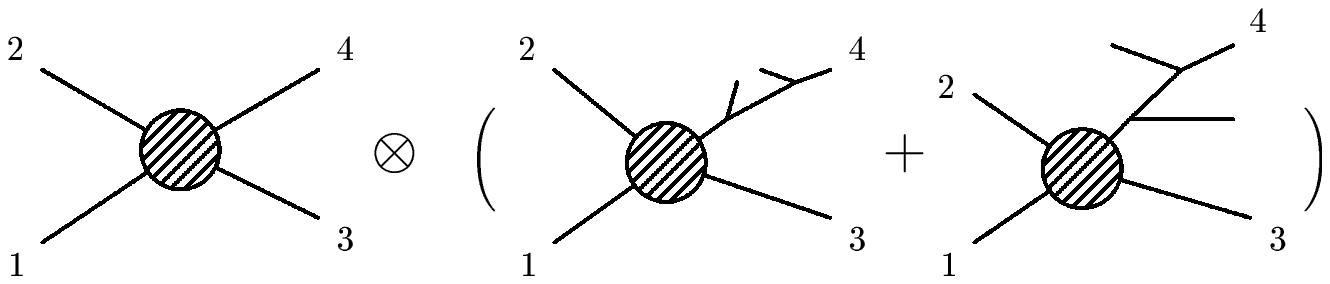, width=12.0cm}
\caption{This figure shows the type I virtual correction on gluon 
4 given by the expression in \eref{eq:v1-4-2}.} 
\label{f:22If} 
\efi

We start with the contribution that has a correction on gluon 4
of the 2-to-2 scattering. Very loosely this can be represented by 
\bea C^{\text{I}(2)4}_{12\lra 34}(p_1) & \simeq & \ing \int [dp_2] 
     d\F_{34} D_{12,34} \frac{1}{2!} \sum \frac{1}{2} 
          \Big ( \cm_{12\ra 34}   \otimes \cm_{12\ra 34}^{\text{I}(2)4*}
                +\cm_{12\ra 34}^* \otimes \cm_{12\ra 34}^{\text{I}(2)4 } 
          \Big )                                                 \nonum \\  
     && \times \Big [ f_1 f_2 (1+f_3)(1+f_4)
                    -(1+f_1)(1+f_2) f_3 f_4 \Big ] F_{\text{I}}(f_4, \dots)
         \; .                                                    \nonum \\  
\label{eq:v1-4}  
\eea  
Here $F_{\text{I}}(f_4, \dots)$ to be given below carries additional 
particle distributions that arise because of the virtual correction 
and is a function of $f_4$ and other internal variables of the 
second order amplitude $\cm^{(2)}$ \cite{wong3}. The amplitude 
$\cm_{12\ra 34}^{\text{I}(2)4}$ carries the type I $\a_s$
correction attached to gluon 4. From the previous subsection we 
already know the form of the correction in the vacuum. Careful
examination of the collinear limit of the virtual contribution shows
that the structure surrounding the pole that gives rise to the 
collinear divergence in the vacuum is the same as the type I part
of the medium contribution that gives rise to the same divergence.
The only difference is whereas the vacuum part has a distribution
of $1$, the medium part has $f$. The only change is then in the 
appearance of a factor $F(f_4, \dots)$ in the integrand of the 
integration over the phase space of the collinear gluons. 
The complete contribution is 
\bea C^{\text{I}(2)4}_{12\lra 34}(p_1) & \simeq & \ing \int [dp_2] 
     d\F_{34} D_{12,34} \frac{1}{2!} \sum |\cm_{12\ra 34}|^2  
     \times \frac{1}{2} \int dz \frac{ds}{s}  \frac{\a_s(s)}{2\p}  
     P_{gg}(z)                                                   \nonum \\  
     && \times \Big [ f_1 f_2 (1+f_3)(1+f_4)
                    -(1+f_1)(1+f_2) f_3 f_4 \Big ] 
               \Big (1+f_4(z)+f_4(1-z) \Big ) \; .  
                                                                 \nonum \\  
\label{eq:v1-4-2}  
\eea  
One can see in \eref{eq:v1-4-2} that there are two different factors 
that are made up of particle distributions. The first one in square 
brackets is the familiar difference of product of distributions that 
ensures this contribution to the collision terms will vanish in 
equilibrium. The second one in parenthesis \cite{wip} 
\be F_{\text{I}}(f_4,z) = 1+f_4(z)+f_4(1-z) \;, 
\ee
is a factor that comes from the virtual correction that is attached 
to gluon 4. They arise when an internal momentum loop has been 
converted into emission and absorption of gluon of the same momentum. 
For this type of correction, it is made up of the emission from and 
then followed by the absorption of gluons of the same momentum by 
gluon 4 \cite{wong3}. This is depicted in \fref{f:22If} where the 
unlabelled lines represent the emitted and then absorbed gluons. The 
form of $F(f_4,z)$ is deduced from detailed examination of this type I 
virtual correction in a medium which is done elsewhere \cite{wip}. 
Although no detail is given here, the correctness of this expression 
can be ascertained in the next section. 

In case that the virtual correction is on gluon 2, it follows 
similarly that 
\bea C^{\text{I}(2)2}_{12\lra 34}(p_1) & \simeq & \ing \int [dp_2] d\F_{34}  
     D_{12,34} \frac{1}{2} \sum |\cm_{12\ra 34}|^2  
     \times \frac{1}{2} \int dz \frac{ds}{s}  \frac{\a_s(s)}{2\p}  
     P_{gg}(z)                                                   \nonum \\  
     && \times \Big [ f_1 f_2 (1+f_3)(1+f_4)
                    -(1+f_1)(1+f_2) f_3 f_4 \Big ] 
               \Big (1+f_2(z)+f_2(1-z) \Big ) \; .  
                                                                 \nonum \\  
\label{eq:v1-2}  
\eea

%%%%%%%%%%%%%%%%%%%%%%%%%%%%%%%%%%%%%%%%%%%%%%%%%%%%%%%%%%%%%%%%%%%%%%%% 
\subsubsection{Type II: virtual correction without vacuum contribution}
\label{ss:vc2} 
%%%%%%%%%%%%%%%%%%%%%%%%%%%%%%%%%%%%%%%%%%%%%%%%%%%%%%%%%%%%%%%%%%%%%%%% 

For the type II virtual corrections, these have no vacuum counterpart, 
when attached to gluon 4, we have again loosely 
\bea C^{\text{II}(2)4}_{12\lra 34}(p_1) & \simeq & \ing \int [dp_2] 
     d\F_{34} D_{12,34} \frac{1}{2!} \sum \frac{1}{2} 
        \Big ( \cm_{12\ra 34}   \otimes \cm_{12\ra 34}^{\text{II}(2)4*}
              +\cm_{12\ra 34}^* \otimes \cm_{12\ra 34}^{\text{II}(2)4 } \Big )
                                                                   \nonum \\ 
     & & \times \Big [ f_1 f_2 (1+f_3)(1+f_4)-(1+f_1)(1+f_2) f_3 f_4 \Big ]
         F_{\text{II}}(f_4, \dots) \;.                                         
\label{eq:v2-4} 
\eea  
Here similar to the same in the previous subsection $F_{\text{II}}$ is 
a function of $f_4$ and internal variables of the second order amplitude 
$\cm^{\text{II}(2)4}_{12\ra 34}$. Again the gluon splitting structure is 
the same independent of whether there is a medium or not, the only 
change when there is a medium is the factor of distribution 
$F_{\text{II}}$. After examining the structure of the virtual 
contribution, one can deduce \cite{wip} 
\bea C^{\text{II}(2)4}_{12\lra 34}(p_1) & \simeq & \ing \int [dp_2] 
     d\F_{34} D_{12,34} \frac{1}{2!} \sum |\cm_{12\ra 34}|^2  
         \times \int dx \frac{dt}{t} \frac{\a_s(|t|)}{2\p}  
         P_{gg}(x)                                                 \nonum \\ 
     & & \times \Big [ f_1 f_2 (1+f_3)(1+f_4)-(1+f_1)(1+f_2) f_3 f_4 \Big ]
                \Big ( f_4(\mbox{$1-x \over x$})
                      -f_4(\mbox{$1 \over x$})   \Big )  \;.       
\label{eq:v2-4-2} 
\eea  
The second factor of distribution in this case is
\be F_{\text{II}}(f_4,x) = f_4(\mbox{$1-x \over x$})
                          -f_4(\mbox{$1\over x$})
\ee
which is a difference of two distributions because there is an
absorption before an emission of a gluon of the same momentum. 
This is why this process can only occur in the medium. (See 
\fref{f:22IIi} for a graphic representation when the correction
is attached to gluon 2. The unlabelled lines on the right are the 
absorbed and then emitted gluons. The freely floating line on the
left is a spectator particle for the interference to make sense.) 
The factor $F_{\text{II}}(f_4,x)$ is again somewhat 
mysterious but similar to the previous subsection, it comes 
from integrating out the internal loop from thermal field theory 
\cite{wong3}. This and the other results are deduced elsewhere 
\cite{wip}. Even though we skipped a lot of steps to arrive at 
\eref{eq:v2-4-2}, one will see in the next section that the expressions 
in this subsection are correct. 

\bfi
\epsfig{file=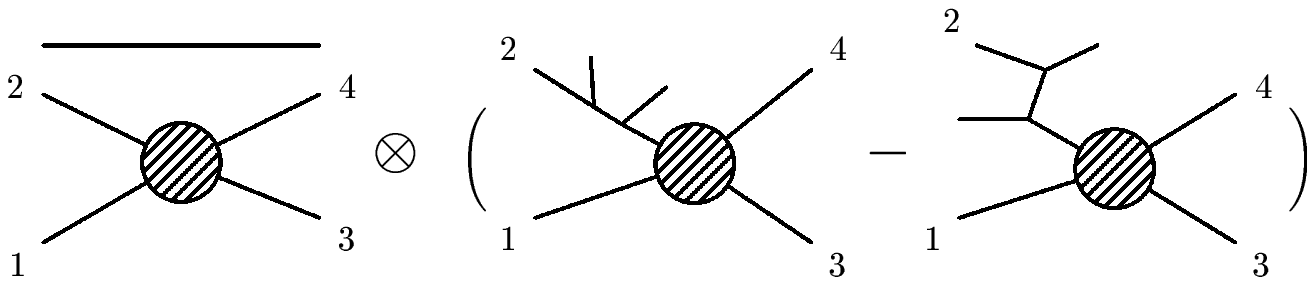, width=12.0cm}
\caption{This figure shows the type II virtual correction on gluon 
2 and represents \eref{eq:v2-2}. On the left hand side, there is
a spectator particle in the $12 \ra 34$ scattering.} 
\label{f:22IIi} 
\efi

In case that the correction is attached to gluon 2, a similar expression
can be derived in the same fashion 
\bea C^{\text{II}(2)2}_{12\lra 34}(p_1) & \simeq & \ing \int [dp_2] 
     d\F_{34} D_{12,34} \frac{1}{2!} \sum |\cm_{12\ra 34}|^2 
         \times \int dx \frac{dt}{t} \frac{\a_s(|t|)}{2\p}  
         P_{gg}(x)                                                 \nonum \\ 
     & & \times \Big [ f_1 f_2 (1+f_3)(1+f_4)-(1+f_1)(1+f_2) f_3 f_4 \Big ]
                \Big ( f_2(\mbox{$1-x \over x$})
                      -f_2(\mbox{$1 \over x$})   \Big )  \;.       \nonum \\  
\label{eq:v2-2} 
\eea  
The graphs associated with this expression is shown in \fref{f:22IIi}.

%%%%%%%%%%%%%%%%%%%%%%%%%%%%%%%%%%%%%%%%%%%%%%%%%%%%%%%%%%%%%%%%%%%%%%%%%%%%%%%
\section{Cancellation of the collinear singularity between the 
real and virtual contributions?}
\label{s:cancel}
%%%%%%%%%%%%%%%%%%%%%%%%%%%%%%%%%%%%%%%%%%%%%%%%%%%%%%%%%%%%%%%%%%%%%%%%%%%%%%%

We will now demonstrate that screening in an out-of-equilibrium 
medium is very important and more so than a plasma in equilibrium. 
This is because collinear singularities cancel in the latter but not 
the former \cite{wip}. Using the already derived results at the 
next-to-leading order, we split up the collision terms into the 
forward and backward process, for example  
\be C_{12\lra 34} = C_{12\ra 34} - C_{34\ra 12}  \;. 
\ee 

%%%%%%%%%%%%%%%%%%%%%%%%%%%%%%%%%%%%%%%%%%%%%%%%%%%%%%%%%%%%%%%%%%%%%%%%%%%%%%%
\subsection{Both collinear gluons are in the final or initial state}
%%%%%%%%%%%%%%%%%%%%%%%%%%%%%%%%%%%%%%%%%%%%%%%%%%%%%%%%%%%%%%%%%%%%%%%%%%%%%%%

From Sec. \ref{ss:bif} for real emission from final state particles, 
one gathers all the possibilities which are the three contributions: 
$3 \pll 4$, $3 \pll 5$, and $4\pll 5$. After relabelling they are 
\be C^{3\pll 4}_{12\ra 345}(p_1)+C^{3\pll 5}_{12\ra 345}(p_1) 
   +C^{4\pll 5}_{12\ra 345}(p_1) = 3\, C^{4\pll 5}_{12\ra 345}(p_1)  \;. 
\ee 
Explicitly this is 
\bea 3\, C^{4\pll 5}_{12\ra 345}(p_1) & \simeq & -\ing \int [dp_2] d\F_{34}  
     D_{12,34} \sum |\cm_{12\ra 34}|^2 \times  \frac{1}{2!} 
     \int dz \frac{ds_{45}}{s_{45}}  \frac{\a_s(s_{45})}{2\p}  
     P_{gg}(z)                                                   \nonum \\  
     && \times f_1 f_2 (1+f_3)(1+f_4(z))(1+f_4(1-z))  \;. 
\label{eq:can-bif}
\eea
In the vacuum (let's assume confinement has been turned off) 
one duplicates this same process by having gluon 1 colliding with 
gluon 2. The collinear divergence here must be cancelled by 
a virtual correction that has a vacuum contribution. Clearly
the results from Sec. \ref{ss:vc1} are required. Summing 
the contributions with virtual correction on gluon 3 and gluon 4,
after relabelling we have  
\be C^{\text{I}(2)3}_{12\ra 34}+C^{\text{I}(2)4}_{12\ra 34} 
   = 2\, C^{\text{I}(2)4}_{12\ra 34} \;,
\ee
where
\bea 2\, C^{\text{I}(2)4}_{12\ra 34}(p_1) & \simeq & \ing 
     \int [dp_2] d\F_{34} D_{12,34} \sum |\cm_{12\ra 34}|^2  
     \times \frac{1}{2} \int dz \frac{ds}{s}  \frac{\a_s(s)}{2\p}  
     P_{gg}(z)                                                   \nonum \\  
     && \times f_1 f_2 (1+f_3)(1+f_4) \Big (1+f_4(z)+f_4(1-z) \Big ) \; .  
\label{eq:v1_12->34}
\eea  
Comparing \eref{eq:can-bif} with \eref{eq:v1_12->34}, the only
difference is in the factor of the particle distributions. That is 
\be \tri f^{\text{I}}(p_4) = (1+f_4(z))(1+f_4(1-z))
                            -(1+f_4) \{1+f_4(z)+f_4(1-z) \} 
\label{eq:df} 
\ee 
\begin{itemize}

\item[(i)]{The plasma is in equilibrium: $f$ takes the form of the 
Bose-Einstein distribution $f_\tBE$ and $\tri f^{\text{I}}(p_4)= 0$ 
is an identity, a consequence of the form of $f_\tBE$. 
This is the cancellation of collinear singularity that we mentioned 
in the introduction. This cancellation occurs even though these 
singularities are screened. A equilibrium system is doubly guarded 
against this type of singularity.
}

\item[(ii)]{The plasma is out of equilibrium: $f$ can take on quite
general forms and is only restricted by requirements such as the
energy of the system must be finite etc. Therefore in general 
the above identity does not hold and $\tri f^{\text{I}}(p_4) \neq 0$.
It follows that there is no longer any cancellation of the 
collinear singularity but since there is the failsafe mechanism
of screening, there is no divergence even in the absence of any
cancellation. 
}

\end{itemize} 

In the case that both collinear gluons are in the initial state,  
the collinear absorption $3\, C^{4\pll 5}_{345\ra 12}(p_1)$ and the 
associated virtual correction $2\, C^{\text{II}(2)4}_{34\ra 12}(p_1)$ 
to the 2-to-2 scattering again differ only by 
\be \tri f^{\text{I}}(p_4) = f_4(z) f_4(1-z)-f_4 \{1+f_4(z)+f_4(1-z) \} \;. 
\ee
After cancelling some of the terms in \eref{eq:df}, this is exactly 
the same $\tri f^{\text{I}}(p_4)$ factor above. The same discussion 
above clearly applies to collinear absorption too.

%%%%%%%%%%%%%%%%%%%%%%%%%%%%%%%%%%%%%%%%%%%%%%%%%%%%%%%%%%%%%%%%%%%%%%%%%%%%%%%
\subsection{One collinear gluon in each of the final and initial state}
%%%%%%%%%%%%%%%%%%%%%%%%%%%%%%%%%%%%%%%%%%%%%%%%%%%%%%%%%%%%%%%%%%%%%%%%%%%%%%%

In Sec. \ref{ss:oif} we saw that in the $12\ra 345$ 5-gluon process 
there are the possibilities of collinear gluon pairs $2 \pll 3$, 
$2 \pll 4$ and $2 \pll 5$. Collecting them together and relabelling, 
we have  
\be C^{2\pll 3}_{12\ra 345}(p_1)+C^{2\pll 4}_{12\ra 345}(p_1)
   +C^{2\pll 5}_{12\ra 345}(p_1) = 3\, C^{2\pll 5}_{12\ra 345}(p_1) \;.
\ee
Explicitly this is 
\bea 3\, C^{2\pll 5}_{12\ra 345}(p_1) & \simeq & -\ing 
     \int [dp_2] d\F_{34} D_{12,34} \frac{1}{2!} \sum |\cm_{12\ra 34}|^2   
         \times \int dx \frac{dt_{25}}{t_{25}} \frac{\a_s(|t_{25}|)}{2\p}  
         P_{gg}(x)                                                 \nonum \\ 
     & & \times f_1 f_2(\mbox{$1 \over x$}) (1+f_3)(1+f_4)  
               (1+f_2(\mbox{$1-x \over x$}))  \;. 
\label{eq:can-oif} 
\eea  
In the vacuum the collinear divergence is cancelled after degenerate
initial states are summed over. It does require other incoming gluons
beside gluon 1 and 2 for that to happen. From the second type of 
virtual correction on gluon 2 to the 2-to-2 scattering in 
Sec. \ref{ss:vc2} 
\bea C^{\text{II}(2)2}_{12\ra 34}(p_1) & \simeq & \ing \int [dp_2] 
     d\F_{34} D_{12,34} \frac{1}{2!} \sum |\cm_{12\ra 34}|^2 
         \times \int dx \frac{dt}{t} \frac{\a_s(|t|)}{2\p}  
         P_{gg}(x)                                                 \nonum \\ 
     & & \times f_1 f_2 (1+f_3)(1+f_4) 
         \Big ( f_2(\mbox{$1-x \over x$})-f_2(\mbox{$1 \over x$}) \Big ) \;.  
\label{eq:v2_12->34}  
\eea  
Comparing \eref{eq:can-oif} and \eref{eq:v2_12->34}, the difference
is entirely in the factor 
\be \tri f^{\text{II}} (p_2) = 
     f_2(\mbox{$1 \over x$}) (1+f_2(\mbox{$1-x \over x$}))
    -f_2 [f_2(\mbox{$1-x \over x$})-f_2(\mbox{$1 \over x$})]  \;.
\label{eq:dfII}
\ee 
Once again we consider the effect of the state of the environment 
has on the interactions. 
\begin{itemize}

\item[(i)]{The plasma is in equilibrium: 
$\tri f^{\text{II}} (p_2) = 0$ is another identity when $f$ 
is the Bose-Einstein distribution $f_\tBE$. This identity ensures the
cancellation of any collinear logarithms between the real and
virtual contributions.}

\item[(ii)]{The plasma is out of equilibrium: the fact that 
the distribution can take many forms means that 
$\tri f^{\text{II}} (p_2) \neq 0$ in general. However any
hint of a divergence is rendered safe by having $m_t^2$ in the 
medium. 
}

\end{itemize} 

Just as a gluon can be emitted from an initial gluon, so a final state
gluon can also absorb an incoming one. Proceeding in a similar fashion
as before, we find the difference between 
$3\, C^{2\pll 5}_{345\ra 12}(p_1)$ and 
$C^{\text{II}(2)2}_{34\ra 12}(p_1)$ is encapsulated in the factor 
\be \tri {f^{\text{II}}}'(p_2) = 
     (1+f_2(\mbox{$1 \over x$})) f_2(\mbox{$1-x \over x$})
    -(1+f_2) [f_2(\mbox{$1-x \over x$})-f_2(\mbox{$1 \over x$})] 
    = \tri f^{\text{II}} (p_2) \;.
\ee 
This is the same expression as in \eref{eq:dfII} and everything 
discussed above evidently also applies.

%%%%%%%%%%%%%%%%%%%%%%%%%%%%%%%%%%%%%%%%%%%%%%%%%%%%%%%%%%%%%%%%%%%%%%%%%%%%%%%
\section{Implications}
\label{s:impl}
%%%%%%%%%%%%%%%%%%%%%%%%%%%%%%%%%%%%%%%%%%%%%%%%%%%%%%%%%%%%%%%%%%%%%%%%%%%%%%%

In the vacuum theory it is well-known that total cross-section
is free of any collinear logarithms. This is true in an equilibrium
system as well when $f=f_\tBE$ as shown in the previous sections. The
special form of $f_\tBE$ ensures that the collinear logarithms
are all cancelled. In neither of these cases have collinear processes 
any influence on the cross-section and the interaction rate is 
unaffected. However this paper is mainly concerned with a system 
that is not yet thermalized, as seen earlier the collinear logarithms 
remain even when the real and virtual processes are combined. To better 
appreciate what this means, let us consider the leading order
2-to-2 process together with only higher order collinear processes. 
In the vacuum and in an equilibrium system, the cross-section will only 
receive contribution from the 2-to-2 process in this special case. 
As soon as the system goes out of equilibrium, the cross-section 
will start receiving contributions at any order from the remaining
collinear processes. There is a huge potential of significant changes 
in the cross-section or the gluon interaction rate in an 
out-of-equilibrium plasma especially when there is a large momentum 
transfer and that the transverse mass $m_t$ is small enough so that 
approximately $\a_s \ln (Q^2/m_t^2) \sim 1$. Whether these collinear
processes can help with thermalization is yet to be seen. However from
the previous sections, it is clear that this depends largely on 
whether each pair of real and virtual collinear process summed up
to give a positive or negative contribution to the collision terms. 
It was shown that the sum of each pair was controlled largely by the 
factors $\tri f^a(p)$ where $a=$ I or II. The signs of these are in 
turned determined by the particle distributions. It is the state of 
the system that has the ultimate control: 
\begin{itemize}

\item[(i)]{it enables the collinear processes to contribute 
to the collision rate,} 

\item[(ii)]{it decides whether these collinear processes will help
or hinder the thermalization of the system.} 

\end{itemize} 

The main question to be answered is if these collinear processes, 
which under normal circumstances would have no effect on the scattering
rate, can have a helping hand in the equilibration of the system. 
From the previous section, the collinear contributions remain 
even after the sum of each real and virtual pair because of 
$\tri f^a(p) \neq 0$. Whether $\tri f^a(p)>0$ or $\tri f^a(p)<0$ will 
help determine largely how these non-vanishing contributions would have 
a positive or negative effect on the collisions (see below). If the 
equilibration of different momentum regions is completely random, 
then $\tri f^a(p)>0$ in one region and $\tri f^a(p)<0$ in another 
there would be no significant net effect. Is there any reason for 
us to expect something different?  

In ref. \cite{betal} the equilibration of a gluon plasma in heavy
ion collisions was described as a perturbative process powered by hard 
gluons. They function as an energy reservoir which pours energy into
the softer momentum regions via gluon radiations so that the lower 
energy gluons reach equilibrium first. This process gradually
spreads upward from lower to higher momenta leading to the 
bottom-up thermalization scenario. This is reasonable for two 
reasons: (i) radiating multiple softer gluons is less restricted by 
energy conservation, (ii) the probability of emitting a lower energy 
gluon is bigger than emitting a higher energy one. As a result
lower momentum regions that have an occupany level below the 
equilibrium level will be filled first ahead of any higher 
momentum regions. In this scenario one can expect the particle 
distributions at momenta below the range of the hard gluon reservoir 
to be organized in a way such that the particle occupancy at lower 
momenta is closer to the equilibrium value than at higher momenta up
to just below the range of the reservoir of hard gluons. That is
if we write 
\be \delta f(p) = f(p)-f_\tBE(p)  \;,
\ee
then $0 \le -\delta f(p) < -\delta f(p')$ for any $p < p'$, where $p$ 
and $p'$ are both below the range of the gluons that are powering the 
equilibration process. 

What impact has the bottom-up scenario on the results that we have
shown so far? Before we answer this question, let us remind ourselves 
what equilibration is. Obviously there are a number of ways that one
can describe it, for example entropy production, momentum transfer
between different momentum regions etc. Here we use the picture of 
particle occupancy in momentum space. Equilibration in this picture
involves reducing the overoccupied momentum regions with respect
to the equilibrium configuration by converting those gluons into 
gluons with momenta in the underpopulated areas. In heavy ion
collisions one can think of the initial conditions as clusters
of hard gluons that overpopulate hard momentum regions mostly aligned 
in the beam directions. Any other directions in momentum space 
would be underoccupied in this case. Alternately if the picture
of the color glass condensates is used instead, the gluons would
overpopulate the transverse plane in the central region and
underpopulate anywhere else \cite{cgc}. In both cases equilibration 
can be described as a process of transferring hard gluons via 
scattering and radiation from one direction or directions on a 
particular plane, which are overpopulated, to vastly different 
directions in other parts of momentum space that are underpopulated. 

Let us gather some important previously mentioned facts and prepare 
them for the central conclusion of this paper. These are the following: 
\begin{itemize}

\item[(i)]{The possibility of equilibration via perturbative QCD 
is our starting premise for the simple reason that this has been 
well tested, better understood and also advocated in \cite{betal}. 
}

\item[(ii)]{The most important contributions for equilibration 
are considered to be large angle collisions because 
they are efficient for transferring momenta between regions in 
momentum space. Yet by themselves they are not very probable so
every hard scattering should be augumented by small angle gluon
radiations and/or absorptions. This augumentation is via the 
compensation of the small coupling by the large collinear logarithm 
when $\a_s \ln(Q^2/m_t^2)$ is of the order of unity. The large
angle scattering thus acts to direct where the gluon radiations
should go but it is the latter that actually fill out the underoccupied
momentum regions. 
}

\item[(iii)]{Because each of the collision terms is made up of 
the difference between the forward and the backward reactions, 
the most helpful processes in the collision terms for achieving 
equilibrium are those that can maximize this difference. There 
are two ways for this to happen. One is when all incoming gluons 
are from the overpopulated momentum regions and all outgoing ones are
in the underoccupied ones. The other is having one incoming flux of 
gluons from the overpopulated region, the other incoming from an 
underoccupied region and all outgoing gluons are in the underoccupied 
ones. The underoccupied region that provides one of the incoming
flux must be from a lower momentum region than the same from the 
overpopulated area. This is so that the higher occupancy of this
lower momentum region can compensate partially for the larger
backward reaction in this case. We will concentrate on scatterings 
that fulfill either condition.    
} 

\item[(iv)]{In the bottom-up equilibration scenario, the particle
distributions in the underpopulated regions satisfy 
\be
     -\delta f(p') > -\delta f(p) > 0 \;, 
\label{eq:df_rel} 
\ee
for any $p < p'$. This implies that
\bea \tri f^{\text{ I}}(p) &=& (1+f(k)) (1+f(p-k))
                              -(1+f(p)) \Big (1+f(k)+f(p-k) \Big ) 
                               \nonum \\ 
                           &=& f(k) f(p-k)
                              -f(p) \Big (1+f(k)+f(p-k) \Big ) 
                               \nonum \\ 
                           &>& 0  \;, 
\label{eq:df-1_rel}
\eea 
and 
\bea \tri f^{\text{II}}(p-k) &=& (1+f(k)) f(p)
                                -f(p-k) \Big (f(k)-f(p) \Big )
                                 \nonum \\
                             &=& (1+f(p)) f(k)
                                -(1+f(p-k)) \Big (f(k)-f(p) \Big )
                                 \nonum \\
                             &=&-\Big \{f(p-k) f(k)
                                - f(p) \Big (1+f(k)+f(p-k) \Big ) 
                                 \Big \} 
                                 \nonum \\ 
                             &=&-\tri f^{\text{ I}} (p) 
                                 \nonum \\
                             &<& 0  \;.     
\label{eq:df-2_rel}
\eea
The simplest way to see that these relations are true is to 
consider a special case. We let $f(k) = f_\tBE(k)$, 
$f(p-k) = f_\tBE(p-k)$ and  $f(p) = f_\tBE(p)+\delta f(p)$ 
where $\delta f(p) < 0$, so
\be \tri f^{\text{ I}}(p) = -\delta f(p) (1+f_\tBE(k)+f_\tBE(p-k)) 
                          = -\tri f^{\text{II}} (p-k) >  0 \;.   
\ee 
For more general distributions that satisfy \eref{eq:df_rel}, the 
relations \etwref{eq:df-1_rel}{eq:df-2_rel} can be more readily
verified numerically. 
}

\item[(v)]{In the overpopulated region where the hard gluons 
provide the energy for the equilibration, the lower energy
gluons here would be closer to equilibrium than the higher energy
ones so we expect for most of these hard gluons
\be \delta f(p') > \delta f(p) > 0 
\ee 
for $p < p'$. At the tail of the distribution at very high energies, 
this is no longer true but the occupancy here becomes too low to  
have too much significance. The consequence is that the relations 
of the $\tri f$'s in (iv) are reversed in the hard gluon reservoir 
\be \tri f^{\text{II}}(p-k) = -\tri f^{\text{ I}}(p) > 0  \;. 
\ee
}

\end{itemize}
%\newpage 

With these points in mind, we now apply them to the gluon interactions.
Without loss of generality, we let $p_1$ to be in the hard gluon
reservoir which supplies the energy for the equilibration. 
Up to the next-to-leading order, the collision terms with the inclusion 
of only collinear processes at and up to order $\a_s^3$ are 
\bea C(p_1) &=& \hspace{0.55cm}  
       C_{12\lra 34}(p_1)
    +\Big (
     3 C^{4\pll 5}_{12\lra 345}(p_1) +2 C^{\text{I}(2)4}_{12\lra 34}(p_1) 
     \Big ) 
    +\Big (
     4 C^{3\pll 5}_{123\lra 45}(p_1) +2 C^{\text{II}(2)4}_{12\lra 34}(p_1)
     \Big )                                         \nonum \\
     & & 
    +2 C^{1\pll 3}_{123\lra 45}(p_1) 
    +\Big (
     3 C^{2\pll 5}_{12\lra 345}(p_1) +  C^{\text{II}(2)2}_{12\lra 34}(p_1)
     \Big )
    +\Big (
       C^{2\pll 3}_{123\lra 45}(p_1) +  C^{\text{I}(2)2}_{12\lra 34}(p_1)
     \Big )                                         \nonum \\  
     & &  
    +2 C^{1\pll 5}_{123\lra 45}(p_1) 
    +3 C^{1\pll 5}_{12\lra 345}(p_1) +  C^{\text{II}(2)1}_{12\lra 34}(p_1)
    +C^{\text{I}(2)1}_{12\lra 34}(p_1) +{\cal O}(\a_s^3)    \;.            
\eea  
Those terms in parenthesis are the pairs of real and virtual 
contributions. Explicitly they are 
\bea & & 3 C^{4\pll 5}_{12\lra 345}(p_1) +2 C^{\text{I}(2)4}_{12\lra 34}(p_1)
                                                                 \nonum \\ 
     &\simeq & 
      -\ing \int [dp_2] d\F_{34} D_{12,34} \sum |\cm_{12\ra 34}|^2  
       \times \frac{1}{2} \int dz \frac{ds}{s}  \frac{\a_s(s)}{2\p}  
     P_{gg}(z)                                                   \nonum \\  
     && \times \Big [ f_1 f_2 (1+f_3)-(1+f_1)(1+f_2) f_3 \Big ] 
        \; \tri f^{\text{I}}(p_4) \; ,     
\label{eq:C_oe}                  
\eea  
\bea & & 4 C^{3\pll 5}_{123\lra 45}(p_1) +2 C^{\text{II}(2)4}_{12\lra 34}(p_1)
                                                                 \nonum \\ 
     &\simeq & 
      -\ing \int [dp_2] d\F_{34} D_{12,34} \sum |\cm_{12\ra 34}|^2 
       \times \int dx \frac{dt}{t} \frac{\a_s(|t|)}{2\p}  
        P_{gg}(x)                                                \nonum \\  
     & & \times \Big [ f_1 f_2 (1+f_3)   
                      -(1+f_1)(1+f_2) f_3 \Big ]
         \tri f^{\text{II}}(p_4)  \; , 
\label{eq:C_oa}
\eea 
\bea & & 3 C^{2\pll 5}_{12\lra 345}(p_1) +C^{\text{II}(2)2}_{12\lra 34}(p_1)
                                                               \nonum \\
     &\simeq & -\ing \int [dp_2] d\F_{34}  
     D_{12,34} \frac{1}{2} \sum |\cm_{12\ra 34}|^2                     
         \times \int dx \frac{dt}{t} \frac{\a_s(|t|)}{2\p}  
         P_{gg}(x)                                             \nonum \\ 
     & & \times \Big [ f_1 (1+f_3)(1+f_4)-(1+f_1) f_3 f_4 \Big ]    
         \tri f^{\text{II}}(p_2)  \; , 
\label{eq:C_ie}                                          
\eea 
and 
\bea & & C^{2\pll 3}_{123\lra 45}(p_1) +C^{\text{I}(2)2}_{12\lra 34}(p_1) 
                                                                 \nonum \\  
     &\simeq & -\ing \int [dp_2] d\F_{34}  
     D_{12,34} \frac{1}{2} \sum |\cm_{12\ra 34}|^2   
     \times \frac{1}{2} \int dz \frac{ds}{s}  \frac{\a_s(s)}{2\p}  
     P_{gg}(z)                                                   \nonum \\ 
     && \times \Big [ f_1 (1+f_3)(1+f_4)-(1+f_1) f_3 f_4 \Big ] 
         \tri f^{\text{I}}(p_2) \; .
\label{eq:C_ia}                  
\eea  
Note that each of these expressions carries either $\tri f^{\text{I}}$ 
or $\tri f^{\text{II}}$ as the last factor. These expressions 
are possible because $\tri f^a$ have at least two equivalent forms. One 
is associated with the forward reactions and the other with the backward 
ones but they are in fact equivalent after some algebra as seen in 
\etwref{eq:df-1_rel}{eq:df-2_rel}. As they stand these expressions 
are hard to analyze, we elect to consider the scattering 
cross-section instead. As a result only the forward part of these 
equations will be used below when we try to construct the scattering
cross-section. 
 
%%%%%%%%%%%%%%%%%%%%%%%%%%%%%%%%%%%%%%%%%%%%%%%%%%%%%%%%%%%%%%%%%%%%%
\subsection{The in-medium cross-section} 
%%%%%%%%%%%%%%%%%%%%%%%%%%%%%%%%%%%%%%%%%%%%%%%%%%%%%%%%%%%%%%%%%%%%%

We consider the interactions in the medium initiated by two flux 
of incoming particles. One flux consists of collinear particles 
centered on the momentum $p_1$. By this we mean gluons of momenta 
$p_1$, or collinear gluons with momenta that can sum up to
$p_1$ such as $p_1-p$ and $p$ where $p_1 \pll p$, 
$|\mathbf{p}_1| > |\mathbf{p}|$ etc are included in the flux. The 
second flux is centered on $p_2$. Through the study of the collisions 
produced by these two flux, we can deduce the effect of the collinear 
processes on the collision rate or equivalently the interaction 
cross-section
\be \sigma_{\text{in-medium}}(p_1, p_2) 
  = -\frac{(2\pi)^3}{\n_g} 
     \Big ( 2 p^0_2 \frac{dC^>_{{\cal F}_1{\cal F}_2}(p_1)}{d^3 p_2} 
           +2 p^0_1 \frac{dC^>_{{\cal F}_1{\cal F}_2}(p_2)}{d^3 p_1} 
     \Big ) 
     \frac{1}{{\cal F}_1{\cal F}_2 \; v_r} 
\label{eq:sigma} 
\ee
where ${\cal F}_i$, $i=$ 1,2 are the particle flux and $v_r$ is the
relative velocity between the two flux. The expression of 
$C^>_{{\cal F}_1{\cal F}_2}$ denotes those forward reactions in the
collision terms that have contributions from interactions between 
the $p_1$-centered with the $p_2$-centered flux and the associated
virtual corrections. This definition of the cross-section reduces
to the usual expression when in the vacuum. The explicit expression 
$C^>_{{\cal F}_1{\cal F}_2}$ up to $\a_s^3$ can be extracted from 
\eref{eq:c22}, \eforef{eq:C_oe}{eq:C_ia}{eq:C_oa}{eq:C_ie}
\bea C^>_{{\cal F}_1{\cal F}_2} (p_1) &=& \hspace{0.35cm} 
     \frac{1}{2} C_{12\ra 34}(p_1) 
    +\frac{1}{2} \Big ( 
     3 C^{4\pll 5}_{12\ra 345}(p_1) +2 C^{\text{ I}(2)4}_{12\ra 34}(p_1) 
     \Big )                                            \nonum \\ 
    & & 
    +\frac{1}{2} \Big (
     4 C^{3\pll 5}_{123\ra 45}(p_1) +2 C^{\text{II}(2)4}_{12\ra 34}(p_1)
     \Big )                                            
    +\Big (
     3 C^{2\pll 5}_{12\ra 345}(p_1) +  C^{\text{II}(2)2}_{12\ra 34}(p_1)
     \Big ) \Big |_{\text{modified}}                   \nonum \\ 
    & &
    +\Big (
       C^{2\pll 3}_{123\ra 45}(p_1) +  C^{\text{ I}(2)2}_{12\ra 34}(p_1)
     \Big )                                            \;. 
\label{eq:C^>} 
\eea 
The factor of one-half in the first three terms is to avoid including
those interactions that are basically symmetric in gluon 1 and 2 twice 
in $\sigma_{\text{in-medium}}$. $C^>_{{\cal F}_1{\cal F}_2} (p_2)$ 
has basically the same expression except gluon 1 is interchanged with 
gluon 2. The two expressions $C^>_{{\cal F}_1{\cal F}_2} (p_1)$ and 
$C^>_{{\cal F}_1{\cal F}_2} (p_2)$ together cover all the interactions 
in $C(p_1)$ except $C^{1\pll 5}_{123\ra 45}(p_1)$. This particular
interaction is initiated by gluon 2, 3 and not 1, thus it
does not contribute to $\sigma_{\text{in-medium}}(p_1, p_2)$. 
Note that the expression \eref{eq:C_ie} have to be modified 
first because the incoming momentum in this expression is roughly 
$p_2/x$ instead of $p_2$. After some adjustments including relabelling 
$p_2 \ra p'_2 = x p_2$, gluon 3 and gluon 4 to distinguish them from 
that in \ethref{eq:C_oe}{eq:C_ia}{eq:C_oa} and freezing the value 
of the incoming $p_2$ instead of $p'_2$, it becomes  
\bea & & \Big ( 
     3 C^{2\pll 5}_{12\lra 345}(p_1) +C^{\text{II}(2)2}_{12\lra 34}(p_1)
         \Big ) \Big |_{\text{modified}}                       \nonum \\
     &\simeq & -\ing \int [dp'_2] d\F_{3'4'}  
     D_{12',3'4'} \frac{1}{2} \sum |\cm_{12'\ra 3'4'}|^2                     
         \times \int dx \frac{dt}{t} \frac{\a_s(|t|)}{2\p}  
         P_{gg}(x)                                             \nonum \\ 
     & & \times \Big [ f_1 (1+f_{3'})(1+f_{4'})-(1+f_1) f_{3'} f_{4'} \Big ] 
         \tri f^{\text{II}} (x p_2) \;. 
\label{eq:C_ie-mod}                     
\eea 
The $d/d^3 p_i$ in \eref{eq:sigma} is there to indicate that 
the integration over $d^3 p_i$ in the different $C$'s should be 
removed and the 4-momentum at $p_i$ is frozen at that value 
algrebraically. 

Let us now examine each term beyond the leading order term in 
\eref{eq:C^>}. Here a term at $\a_s^3$ refers to each pair of real and 
virtual contribution enclosed in parenthesis. From the above discussions, 
point (iv) guarantees that the second and third term give a positive 
and negative contribution respectively to $\sigma_{\text{in-medium}}$. 
These two terms differ only by a collinear emission $4\ra ij$ from 
and a collinear absorption $4j\ra i$ by one of the final particles. 
Apart from the $ds$ and $dt$ integrations that yield the slowly 
varying logarithms \etwref{eq:ln1}{eq:ln2} which are both essentially 
controlled by the $Q^2/m^2_t$ ratio, one has the $dz$ integration 
over $\frac{1}{2}\tri f^{\text{I}}(p_4)$ and the other has $dx$ 
over $\tri f^{\text{II}}(p_4)$. While 
$\tri f^{\text{I}}(p_4) > -\tri f^{\text{II}}(p_4) > 0$ for all $p_4$ 
in the underpopulated momentum region, the same cannot be said
of $\frac{1}{2}\tri f^{\text{I}}(p_4) > -\tri f^{\text{II}}(p_4)$
which is not always true. However 
\be       \lim_{z\ra 0} \tri f^{\text{ I}}(p_4) 
    =     \lim_{z\ra 1} \tri f^{\text{ I}}(p_4) 
    \Lra -\lim_{x\ra 1} \tri f^{\text{II}}(p_4) \;.
\ee
While $\tri f^{\text{II}}(p_4)$ decreases monotonically towards
zero as $x$ decreases from unity, $\tri f^{\text{I}}(p_4)$ has a 
minimum at $z = 1/2$ and two maxima at $z=0$ and $z=1$ as can be 
seen in \fref{f:Dfx}. These maxima can be quite large. This leads 
to the result that after integrating over the light-cone fractions 
$x$ and $z$ in the second and third terms of \eref{eq:C^>}, the 
second and positive contributing term will be larger than the 
third negative term (see \fref{f:Dfx}). The sum of the second 
and third terms then contribute positively to 
$\sigma_{\text{in-medium}}$. Of course the limits that should be
used are $\zmn < z < \zmx$ and $\xmn < x < \xmx$ shown in 
Sec. \ref{s:htl} but these are not too far from 0 and 1 when the 
ratio of $Q^2/m_t^2$ is large. 

\bfi
\epsfig{file=Dfx.eps, width=8.0cm}
\caption{A comparison of the modulus of the typical 
$\tri f^{\text{I}}(p)$ and $\tri f^{\text{II}}(p)$ for some fixed 
$p$ in an underpopulated momentum region for various values of the 
lightcone fraction $z$ or $x$. The area covered by the curve of 
$\tri f^{\text{I}}(p)$ is clearly more than twice that of 
$\tri f^{\text{II}}(p)$.} 
\label{f:Dfx}
\efi

There are the remaining fourth and fifth terms. For these we have
to distinguish the two different incoming gluons raised in point (iii). 
\begin{itemize} 

\item[1)]{Both flux from overpopulated momentum regions: 

The fourth term of \eref{eq:C^>} 
has $\tri f^{\text{II}}(x p_2)=-\tri f^{\text{I}}(p_2)>0$
from \eref{eq:C_ie-mod} (point (v)), as part of the integrand and 
the fifth has $\tri f^{\text{I}}(p_2)$ from \eref{eq:C_ia}. Thus 
the fourth has a positive contribution to the cross-section and the 
fifth contributes negatively. Although they both carry the same 
factor $\tri f^{\text{I}}(p_2)$, the fifth term has an extra 
symmetry factor of one-half and the occupancies of the outgoing 
particles are smaller than those from the fourth term because 
the outgoing gluon momenta are in general larger. To see this,
the outgoing momenta of the fourth term are 
$p_{3'}+p_{4'} = p_1+p_2-p_5$ whereas those of the fifth term
are $p_{3}+p_{4} = p_1+p_2$. $f(p_{3'})$ and $f(p_{4'})$ will 
in general be larger than $f(p_{3})$ and $f(p_{4})$ as a 
consequence. The sum of the fourth and fifth term therefore 
gives a net positive contribution to $\sigma_{\text{in-medium}}$. 
Grouping now all the contributions of collinear processes from 
the final states as well as those in the initial states discussed 
here, the sum of the four next-to-leading contributions in 
\eref{eq:C^>} are negative so their contribution to the in-medium 
cross-section is positive. 
} 

\item[2)]{One flux from an overpopulated and the other from an 
underpopulated region: 

In this case the $p_2$-centered flux is from an underpopulated 
region since $p_1$ has already been chosen to be from the hard 
overpopulated region. This is the reverse of situation of 1). From 
point (iv) the fourth term of \eref{eq:C^>} is now negative with 
$-\tri f^{\text{I}}(p_2)<0$ and the fifth becomes positive with 
$\tri f^{\text{I}}(p_2)>0$. The same reasons given in 1) that ensured 
the sum of the two was positive there now ensure that it is negative. 
To sum it up, the second and third term of \eref{eq:C^>} give a
net positive contribution but the fourth and fifth term give a
net negative contribution. 

Of course these are not the only contributions because both 
$C^>_{{\cal F}_1{\cal F}_2} (p_1)$ and 
$C^>_{{\cal F}_1{\cal F}_2} (p_2)$ have an entry in 
$\s_{\text{in-medium}}$ in \eref{eq:sigma}. In the 
$C^>_{{\cal F}_1{\cal F}_2} (p_2)$ entry, the second and third 
term and also the fourth and fifth term as well would still give 
net positive contributions because there is not much difference 
here from case 1). It follows that the collinear processes 
in the initial states of the two incoming flux in this case  
tend to have opposing and compensating effects on each other. 
In general it is hard to tell which contributions will have the
upper-hand since it depends to a certain extent both on the value
of $p_1$ and $p_2$, and on how close each of these is to the boundary 
separating the overpopulated to the under-occupied regions.    
Thus the collinear processes in the initial states do not have 
much net effect on the average, however those in the final states 
are still very much present and positive. One can conclude that 
there is still enhancing effect on $\s_{\text{in-medium}}$ but to a 
lesser degree than that in 1). 
} 

\end{itemize}

%%%%%%%%%%%%%%%%%%%%%%%%%%%%%%%%%%%%%%%%%%%%%%%%%%%%%%%%%%%%%%%%%%%%%
\section{In conclusion}
%%%%%%%%%%%%%%%%%%%%%%%%%%%%%%%%%%%%%%%%%%%%%%%%%%%%%%%%%%%%%%%%%%%%%

Because of the overall positive contributions from the collinear
processes together with the fact that these contributions 
vanish when the interactions occur in the vacuum or when 
the system is in equilibrium, it means that the in-medium 
non-equilibrium cross-section is larger or equivalently 
the out-of-equilibrium interaction rate is higher than what is 
expected. This is the novel perturbative mechanism for achieving 
very fast thermalization within 0.6 fm/c at RHIC that we are hoping
for. Underlying this mechanism is the non-vanishing of collinear 
logarithmic terms in a system that is out of equilibrium. We are 
not aware that this has been pointed out in any existing literature. 
This is considered in more detail in \cite{wip}. In closing we have
yet to consider here the size of the enhancement which is clearly
dependent on how far away from equilibrium the system is, and also 
we have not considered how the contributions at higher orders will 
affect our main conclusion, nevertheless a lot of the structures in 
the various $C(p_i)$'s, the factors of distributions and the inequalities 
that they satisfy are shared with those at higher orders. These are 
better studied and verified numerically than by using the arguments 
that we have attempted here. All these will be done in the future.

%%%%%%%%%%%%%%%%%%%%%%%%%%%%%%%%%%%%%%%%%%%%%%%%%%%%%%%%%%%%%%%%%%%%%%%%%%%%%
\section*{Acknowledgements}
%%%%%%%%%%%%%%%%%%%%%%%%%%%%%%%%%%%%%%%%%%%%%%%%%%%%%%%%%%%%%%%%%%%%%%%%%%%%%

S.W. thanks U. Heinz, A. Mueller, B. M\"uller and M. Gyulassy for valuable 
discussions, and U. Heinz, A. Mueller and B. M\"uller for reading the 
manuscript, E. Braaten for helping with the hard thermal loop section. 
This work was supported in part by the U.S. Department of Energy under 
Contract No. DE-FG02-01ER41190.

\bigskip

%%%%%%%%%%%%%%%%%%%%%%%%%% References %%%%%%%%%%%%%%%%%%%%%%%%%%%%%%%%%%%%%%%%%


\begin{thebibliography}{99}
\bibitem{wip}Work in progress to be reported elsewhere.
\end{thebibliography}
\end{document}